%% file: EXO-12-047_temp.tex
\pdfoutput=1

\documentclass[11pt,twoside,a4paper,cmspaper,final,collab]{cms-tdr}
\def\svnVersion{326256}\def\cmsCernNo{2013-037}\def\cmsCernDate{\today}\def\cmsMessage{Published in Physics Letters B as \href{http://dx.doi.org/10.1016/j.physletb.2016.01.057}{\doi{10.1016/j.physletb.2016.01.057.}}}

\begin{document}\cmsNoteHeader{EXO-12-047}

\hyphenation{had-ron-i-za-tion}
\hyphenation{cal-or-i-me-ter}
\hyphenation{de-vices}
\RCS$HeadURL: svn+ssh://svn.cern.ch/reps/tdr2/papers/EXO-12-047/trunk/EXO-12-047.tex $
\RCS$Id: EXO-12-047.tex 326223 2016-02-17 15:09:04Z gomber $
\newlength\cmsFigWidth
\ifthenelse{\boolean{cms@external}}{\setlength\cmsFigWidth{0.85\columnwidth}}{\setlength\cmsFigWidth{0.4\textwidth}}
\ifthenelse{\boolean{cms@external}}{\providecommand{\cmsLeft}{top}}{\providecommand{\cmsLeft}{left}}
\ifthenelse{\boolean{cms@external}}{\providecommand{\cmsRight}{bottom}}{\providecommand{\cmsRight}{right}}
\ifthenelse{\boolean{cms@external}}{\providecommand{\EmT}{\ensuremath{\widetilde{E}\hspace{-0.55em}/}\xspace}}{\providecommand{\EmT}{\ensuremath{\widetilde{E}\hspace{-0.6em}/}\xspace}}
\newcommand{\AeMC}{\ensuremath{A \epsilon_{\scriptscriptstyle{\mathrm{MC}}}}\xspace}

\cmsNoteHeader{EXO-12-047} 
\title{Search for new phenomena in monophoton final states in proton-proton collisions at $\sqrt{s}=8\TeV$}

\date{\today}

\abstract{Results are presented from a search for new physics in final states containing a photon and missing transverse momentum. The data correspond to an integrated luminosity of 19.6\fbinv collected in proton-proton collisions at $\sqrt{s}=8\TeV$ with the CMS experiment at the LHC. No deviation from the standard model predictions is observed for these final states. New, improved limits are set on dark matter production and on parameters of models with large extra dimensions. In particular, the first limits from the LHC on branon production are found and significantly extend previous limits from LEP and the Tevatron. An upper limit of 14.0\unit{fb} on the cross section is set at the 95\% confidence level for events with a monophoton final state with photon transverse momentum greater than 145\GeV and missing transverse momentum greater than 140\GeV.}

\hypersetup{%
pdfauthor={CMS Collaboration},%
pdftitle={Search for new phenomena in monophoton final states in proton-proton collisions at sqrt(s) = 8 TeV},%
pdfsubject={CMS},%
pdfkeywords={CMS, physics, monophoton, dark matter}}

\maketitle 

\section{Introduction}

The production of events containing photons with large transverse momentum and having large missing transverse momentum at the CERN LHC is sensitive to physics beyond the standard model (SM).
In this Letter we investigate three possible extensions of the SM: a model incorporating pair production of dark matter (DM) particles, and two models with extra spatial
dimensions, as described below.

At the LHC, DM particles ($\chi$)~\cite{DMGeneral} can be produced in the process
$\Pq\Paq \to\Pgg \chi\overline{\chi}$, where the photon is radiated by one of the incoming quarks.
With a photon in the final state, we gain sensitivity to the production of invisible particles.
The SM-DM interaction is assumed to be mediated by a virtual particle (``mediator``) with a mass $M$ much heavier than the fermionic DM particle mass ($M_{\chi}$).
Various processes are contracted into an effective field theory (EFT)~\cite{DMTeva,DMLHC1,Majorana_dm,DMCollid}, assuming $M$ much larger than the momentum transfer scale $Q$ (i.e. $M \gg Q$) and a contact interaction scale $\Lambda$ given by $\Lambda^{-2} = g_{\chi} g_\Pq M^{-2}$, where $g_{\chi}$ and $g_\Pq$ are the mediator couplings to $\chi$ and to quarks, respectively.
Using this formalism, results from searches at the LHC can be related to limits for direct searches sensitive to $\chi$-nucleon scattering~\cite{DMCollid}.

The ADD model~\cite{ADD,ADD1} of large extra dimensions
is postulated to have $n$ extra compactified spatial dimensions at a characteristic scale $R$ that reflects an effective Planck scale
$\MD$ through $M_\text{Pl}^2 \approx \MD^{n+2}R^n$, where $M_\text{Pl}$ is the Planck scale.
If $\MD$ is of the same order as the electroweak scale ($M_\mathrm{EW} \sim 10^{2}$\GeV), the large value of $M_\text{Pl}$ can be interpreted as being a consequence of large-volume (${\sim}R^{n}$) suppression from extra dimensional space.
This model predicts a sizable cross section for the process $\Pq\Paq\to \Pgg\PXXG$, where $\PXXG$ is a graviton that escapes detection, and motivates the search for events
with a single $\gamma$ and missing transverse momentum.

In both the ADD and branon models, the SM particles are constrained to live on a 3+1 dimensional 3-brane surface.
In the branon family of models~\cite{sundrum,dobado,cembranos1,cembranos3}, it is assumed that the brane fluctuates in the extra dimensions, in contrast to the ADD model, where the brane is rigid.
In this alternative scheme, the brane tension scale $f$ is expected to be much
smaller than other relevant scales such as $\MD$. The particles associated with such fluctuations are scalar particles called branons. Branons are stable and massive scalar particles
of mass $M_B$, and are natural candidates for dark matter~\cite{branon_dm}.
 They can be pair-produced in association with SM
 particles at the LHC, giving rise to $\Pgg+$missing transverse momentum final states~\cite{cembranos2}.
If $N$ extra dimensions are considered, then $N$ branons are expected and their production cross section scales with $N$. In the following, only the $N=1$ case is considered.

The primary background to the $\Pgg+$missing transverse momentum signal is the irreducible SM background from $\cPZ\Pgg\to\Pgn\Pagn\Pgg$
 production.
Other backgrounds include $\PW\Pgg\to\ell\Pgn\Pgg$ (where $\ell$ is an undetected charged lepton), $\PW\rightarrow\Pe\Pgn$ ( where the electron is misidentified as a photon), $\Pgg$+jet, QCD multijet (with a jet misidentified as a photon), $\cPZ\Pgg\to\ell\ell\Pgg$, and diphoton events,
as well as backgrounds from beam halo.

\section{The CMS detector}
The CMS experiment uses a right-handed coordinate system, with the origin at the nominal interaction point, the $x$ axis pointing to the center of the LHC, the $y$ axis pointing up (perpendicular to the LHC plane), and the $z$ axis along the anticlockwise-beam direction. The azimuthal angle $\phi$ is measured from the $x$-axis in the $x$-$y$ plane and the polar angle $\theta$ is measured from the $z$-axis. Pseudorapidity is defined as $\eta = -\ln[ \tan(\theta/2)]$.

The central feature of the CMS apparatus is a superconducting solenoid of 6\unit{m} internal diameter, providing a magnetic field of 3.8\unit{T}. Within the superconducting solenoid volume are a silicon pixel and strip tracker, a lead tungstate crystal electromagnetic calorimeter (ECAL), and a brass and scintillator hadron calorimeter (HCAL), each composed of a barrel ($\abs{ \eta }< 1.479$) and two endcap ($1.479 <\abs{ \eta } < 3.0$) sections. Electrons are found by associating clusters of ECAL energy with adjacent tracker hits. Muons are detected in the pseudorapidity range $\abs{\eta}< 2.4$, using gas-ionization detectors embedded in the steel flux-return yoke outside the solenoid, and reconstructed from tracks in these detectors combined with those from the silicon tracker. Extensive forward calorimetry ($3.15 <\abs{ \eta }< 4.9$) complements the coverage provided by the barrel and endcap detectors. The energy resolution for photons with
transverse momentum ${\ge}$ 60\GeV varies between 1.1\% and 2.6\% over the solid angle of the ECAL barrel, and from 2.2\% to 5.0\% in the endcaps~\cite{Chatrchyan:2013dga}. The timing measurement of the ECAL has a resolution better than 200\unit{ps} for energy deposits larger than 10\GeV~\cite{Chatrchyan:2013dga}. In the $\eta$-$\phi$ plane, and for $\abs{\eta}< 1.48$, the HCAL cells map onto $5{\times}5$ arrays of ECAL crystals to form calorimeter towers projecting radially outward from the nominal interaction point. A more detailed description of the CMS detector, together with a definition of the coordinate system used and the relevant kinematic variables, can be found in Ref~\cite{JINST1}.

\section{Event selection}
In the following, it is convenient to refer to the missing transverse momentum vector, $\vec{\ETslash}$, defined as the projection on the plane perpendicular to the beams of the negative vector sum of the momenta of all reconstructed particles in an event. Its magnitude is referred to as \ETslash.

Events are selected from a data sample corresponding to an integrated luminosity of 19.6\fbinv collected in proton-proton collisions at $\sqrt{s}=8$\TeV with the CMS experiment at the LHC. Triggers requiring at least one electromagnetic cluster or a cluster along with large \ETslash are used.
For the selected signal region of transverse energy $\ET^{\gamma}>$ 145\GeV, pseudorapidity $\abs{\eta^{\Pgg}} <1.44$, and $\ETslash > 140$\GeV, these triggers are $\approx$96\%
efficient for $\ET^{\gamma}$ in the 145--160\GeV range, and fully efficient for $\ET^{\gamma} >160$\GeV.
Events are required to have at least one primary vertex reconstructed within a longitudinal distance of $\abs{z} < 24\unit{cm}$  of the center of the detector
and at a distance $<$2\unit{cm} from the $z$-axis. The primary vertex is chosen to be the vertex with the highest sum in $\pt^{2}$ of its associated tracks, where $\pt$ is the transverse momentum.

Candidate electromagnetic (EM) showers are restricted to the barrel region of the ECAL, where their purity is highest~\cite{purity}.
Photon candidates~\cite{CMS-PAS-EGM-10-006} are selected by requiring the ratio of the energy deposited in the closest HCAL tower to
 the energy of the EM showers in the ECAL to be less than 0.05 and the spatial distribution of energy in the EM shower to be
consistent with that expected for a photon.
In order to reject hadronic activity, photon candidates are required to be isolated, using the sum of the transverse energy of additional particles within a cone of $\Delta R < 0.3$ centered on the shower axis, where $\Delta$R = $\sqrt{\smash[b]{(\Delta\eta)^{2} + (\Delta\phi)^{2}}}$, reconstructed using a particle-flow algorithm~\cite{CMS-PAS-PFT-09-001,CMS-PAS-PFT-10-001}.
In this isolation cone, the sum of the transverse energy (in \GeVns{}) of additional photons is required to be less than ($0.7 + 0.005\ET^{\gamma}$), of neutral hadrons is required to be less than ($1.0 + 0.04\ET^{\gamma}$), and of charged
hadrons is required to be less than 1.5. The charged hadron contribution includes that calculated from the other interaction vertices in the event (pileup), arising from the uncertainty in assigning
the photon candidate to a particular vertex. The effect of pileup on the isolation variables is mitigated using the scheme presented in Ref.~\cite{FastJet1}.

The ECAL crystal containing the highest energy within the cluster of the photon candidate is required
to have a time of deposition within $\pm$3\unit{ns} of particles arriving from the
collision. This selection suppresses contributions from noncollision backgrounds. To reduce contamination from beam halo, the crystals (excluding those associated with the photon candidate) are
examined for evidence of the passage of a minimum-ionizing particle roughly parallel to the beam axis (beam halo tag).
If sufficient energy is found along such a trajectory, the event
is rejected. Highly ionizing particles traversing the sensitive volume of
the readout photodiodes can give rise to spurious signals within the EM shower~\cite{Petyt:2012upa}.
These EM showers are eliminated by requiring consistency among the timings
 of energy depositions in all crystals
 within the shower.
Photon candidates are rejected if they are likely to be electrons, as inferred from characteristic patterns of hits in the pixel detector, called ``pixel seeds'',
that are matched to candidate EM showers~\cite{EGM-10-004}.

Jets are reconstructed with the anti-\kt algorithm~\cite{Cacciari:2008gp} using a radius parameter of $R = 0.5$.
Jets that are identified as arising from pileup are rejected~\cite{CMS-PAS-JME-13-005}. In order to reduce QCD multijet backgrounds,
events are rejected if there is more than one jet with $\pt > 30\GeV$ at $\Delta R > 0.5$ relative to the photon.
Events with isolated leptons (electron or muon) with $\pt > 10\GeV$, $\abs{\eta}< 2.4$ ($2.5$) for muons (electrons) and $\Delta R > 0.5$ relative to the photon, are also rejected to suppress $\PW\Pgg\to\ell\Pgn\Pgg$ and $\cPZ\Pgg\to\ell\ell\Pgg$ backgrounds. Lepton isolation is computed using the sum of transverse
energies of tracks, ECAL, and HCAL depositions within a surrounding cone of $\Delta R < 0.3$.
For electron isolation, each contributing component of transverse energy (tracker, ECAL, and HCAL) is required to be less than 20\% of the electron \pt,
while for muons only the tracker component is considered and is required to be less than 10\% of the muon \pt.

The candidate events are required to have $\ETslash > 140\GeV$.
A topological requirement of $\Delta \phi(\vec{\ETslash},\gamma) > 2$\unit{rad} is applied to suppress the contribution from the $\Pgg$+jet background.

A major source of background comes from events with mismeasured \ETslash due to finite detector resolution, mainly associated with jets. In order to reduce the contribution from events with mismeasured \ETslash, for each event a $\chi^{2}$ function is constructed and minimized:
\begin{equation}\label{eq:1a}
   \chi^2 = \sum_{i} \left( \frac{(\PT^\text{reco})_i-(\widetilde{p}_{\mathrm {T}})_i}{(\sigma_{\PT})_i} \right)^2 + \left( \frac{\EmT_x}{\sigma_{\EmT_x}}\right)^2  + \left( \frac{\EmT_y}{\sigma_{\EmT_y}}\right)^2,
\end{equation}
where the summation is over the reconstructed particles, \ie, the photon and the jets. In the above equation, $(\PT^\text{reco})_{i}$ are the transverse  momenta, and the $(\sigma_{\PT})_{i}$, the expected momentum resolutions of the reconstructed particles. The $(\widetilde{p}_\mathrm{T})_{i}$ are the free parameters allowed to vary in order to minimize the function. The resolution parametrization associated with the \ETslash is obtained from Ref~\cite{metreso}. Lastly, $\EmT_x$ and $\EmT_y$ can be expressed as:
\begin{equation}
\begin{split}
\EmT_{x,y} &= \Em^\text{reco}_{x,y} + \sum_{i=\text{objects}} (p_{x,y}^\text{reco})_i - (\widetilde{p}_{x,y})_i  \\
\\
&=  - \sum_{i=\text{objects}} (\widetilde{p}_{x,y})_i
\end{split}
\end{equation}
In events with no genuine \ETslash, the mismeasured quantities will be more readily re-distributed back into the particle momenta, which will result in a low $\chi^{2}$ value. On the other hand, in events with genuine \ETslash from undetected particles, minimization of the $\chi^{2}$ function will be more difficult and generally will result in larger $\chi^{2}$ values. To reduce the contribution of events with mismeasured \ETslash, the probability value obtained from the $\chi^{2}$ minimization is required to be smaller than $10^{-6}$ and $\EmT_\mathrm{T} = \sqrt{\smash[b]{\EmT_{x}^2 + \EmT_{y}^2}}$, in which the original reconstructed particle momenta are replaced with those obtained with the $\chi^{2}$ minimization, is required to be greater than 120\GeV. These requirements are optimized using the significance estimator $\mathrm{S}/\sqrt{\mathrm{S}+\mathrm{B}}$ and remove 80\%\,(35\%) of $\Pgg$+jet (QCD multijet) events, while keeping 99.5\% of signal events.

After applying all selection criteria, 630 candidate events remain in the sample.
\section{Background determination}

Backgrounds from $\cPZ\Pgg\to\Pgn\Pagn\Pgg$, $\PW\Pgg\to\ell\Pgn\Pgg$, $\Pgg$+jet, $\cPZ\Pgg\to\ell\ell\Pgg$, and
diphoton production are estimated from simulated samples processed through the full \GEANTfour-based simulation of the CMS detector \cite{GEANT, GEANTdev}, trigger emulation, and the same event reconstruction programs as used for data. The $\cPZ\Pgg\to\Pgn\Pagn\Pgg$ and  $\PW\Pgg\to\ell\Pgn\Pgg$ samples are generated with
\MADGRAPH5v1.3.30~\cite{Madgraph_new}, and the cross section is corrected to include next-to-leading-order (NLO) effects through an $\ET^{\gamma}$ dependent correction factor calculated with \MCFM6.1~\cite{MCFM}. The central values of the NLO cross section and the prediction for the photon $\ET$ spectrum are calculated following the prescriptions of the PDF4LHC Working Group~\cite{Alekhin:2011sk,PDF4LHC,PDF4LHC1}. This prescription is also used to calculate the systematic uncertainties due to the parton distribution functions (PDF), and the strong coupling $\alpha_{s}$ and its dependence on the factorization scale and renormalization scale. The systematic uncertainties in the NLO cross sections are found to be in the range 8\% to 48\% and 16\% to 82\% for $\cPZ\Pgg\to\Pgn\Pagn\Pgg$ and $\PW\Pgg\to\ell\Pgn\Pgg$, respectively, over the $\ET^{\gamma}$ spectrum from 145\GeV to 1000\GeV. The strong correlation in the uncertainties of the two channels is propagated to the final result.
The $\cPZ\Pgg\to\ell\ell\Pgg$ sample is obtained using the \MADGRAPH5v1.3.30 generator~\cite{Madgraph_new}.
The  $\Pgg$+jet and diphoton samples are obtained using the \PYTHIA6.426 generator~\cite{Pythia6}
 at leading order (LO), with the CTEQ6L1~\cite{CTEQ6L1} PDF. The  $\Pgg$+jet cross section is corrected to include NLO effects.

The backgrounds estimated from simulations are scaled by a factor $F$ to correct for observed differences in efficiency between data and simulation.
This overall data/simulation correction factor receives contributions from four sources as follows: the photon reconstruction efficiency ratio, estimated to be $0.97\pm0.02$
using $\cPZ\to\Pe\Pe$ decays; the ratio of probabilities for satisfying a crystal timing requirement, estimated to be $0.99\pm0.03$ from a sample of electron data; the lepton veto efficiency ratio, estimated to be $0.99\pm0.02$ using $\PW\to\Pe\Pgn$ decays; and the jet veto efficiency ratio, estimated to be $0.99\pm0.05$
using $\PW\to\Pe\Pgn$ decays, and confirmed using $\cPZ\Pgg\to\Pe\Pe\Pgg$ data samples. The total correction factor obtained by combining these contributions is $F = 0.94\pm0.06$.

The total uncertainty in the backgrounds estimated through simulation includes contributions from the theoretical cross section, data-simulation factor $F$, pileup modeling,
and the accuracy of energy calibration and resolution for photons~\cite{Chatrchyan:2013dga}, jets~\cite{JetEnCor2011V2}, and \ETslash~\cite{JME-10-009}.
The estimated contribution from the $\cPZ\Pgg\to\Pgn\Pagn\Pgg$ and $\PW\Pgg\to\ell\Pgn\Pgg$ processes to the background
are, respectively, $345\pm 43$ and $103 \pm 21$ events, where the dominant uncertainty is from the theoretical cross section calculations.
To gain confidence in the estimates from simulation, control regions, which are dominated by these backgrounds and have negligible contributions from a signal, are defined in the data. As a crosscheck, the total contribution from $\cPZ\Pgg\to\Pgn\Pagn\Pgg$ is estimated in data using a sample of
$\cPZ\Pgg\to\mu\mu\Pgg$ candidates, where the muons from the decay of the Z boson are considered as invisible particles hence contributing to \ETslash~\cite{monojet2014}.
The normalization is corrected both for the ratio of the branching fractions of $\cPZ\Pgg\to\Pgn\Pagn\Pgg$ and $\cPZ\Pgg\to\mu\mu\Pgg$, and for differences in the acceptance and selection efficiencies.
This crosscheck provides an estimate of $341 \pm 50$ events, where the uncertainty is dominated by the
 size of the sample.
A control region dominated by the $\PW\gamma$ process is also studied by using the signal selection but inverting the lepton veto \ie, the final state is required to contain a reconstructed charged lepton.
After this selection, 104 events are observed and $126\pm23$ are expected.

Electrons misidentified as photons arise mainly from highly off-shell W boson ($\PW^{*}\to\Pe\Pgn$) events. These backgrounds are inclusively estimated from data.
The efficiency, $\epsilon_\text{pix}$, of matching electron showers in the calorimeter to pixel seeds is estimated using a tag-and-probe technique~\cite{tp}
on $\cPZ\to \Pe\Pe$ events in data, verified with simulated events. The efficiency is found to be $\epsilon_\text{pix}=0.984\pm0.002$ for electrons with $\ET> 100$\GeV. A control sample of $\PW^{*}\to\Pe\Pgn$ events is also obtained from data through use of all the standard candidate selections, with the exception of the pixel seed, which is inverted. The number of events in this sample is scaled by the value of $(1-\epsilon_\text{pix})/\epsilon_\text{pix}$
resulting in an inclusive estimate of $60\pm6$ $\PW^{*}\to\Pe\Pgn$ events in the signal region.

The contamination from jets misidentified as photons is estimated in data using a control sample with $\ETslash < 30\GeV$, dominated by QCD events. This sample is used to measure the ratio of the number of objects that pass photon identification criteria to the number that fail at least one of the isolation requirements.
The control sample also contains objects from QCD direct photon production that must be removed from the numerator
of the ratio. This contribution is estimated by fitting the shower
shape distribution with template distributions. For true photons, a template
for the shower width is formed using simulated $\gamma$+jets events. For jets
misidentified as photons, the template is formed using a separate control
sample, where the objects are required to fail charged
hadron isolation. This corrected ratio is used to scale a set of
data events that pass the denominator selection of the fake ratio and
all other candidate requirements, providing an inclusive estimate for all backgrounds in which jets are misidentified as photons of $45 \pm 14$ events.

Noncollision backgrounds are estimated from data by examining the
shower width of the EM cluster and the
time-of-arrival of the signal in the crystal containing the largest deposition of energy.
Templates for anomalous signals, cosmic ray muons, and beam halo events are obtained by inverting the shower shape and beam halo tag requirements, and are fitted to the timing distribution
of the candidate sample.
The only nonnegligible residual contribution to the candidate sample is found to arise from the beam halo, with an estimated $25\pm6$ events.

\section{Results}
\begin{figure}[!htb]
\centering
\includegraphics[width=0.48\textwidth]{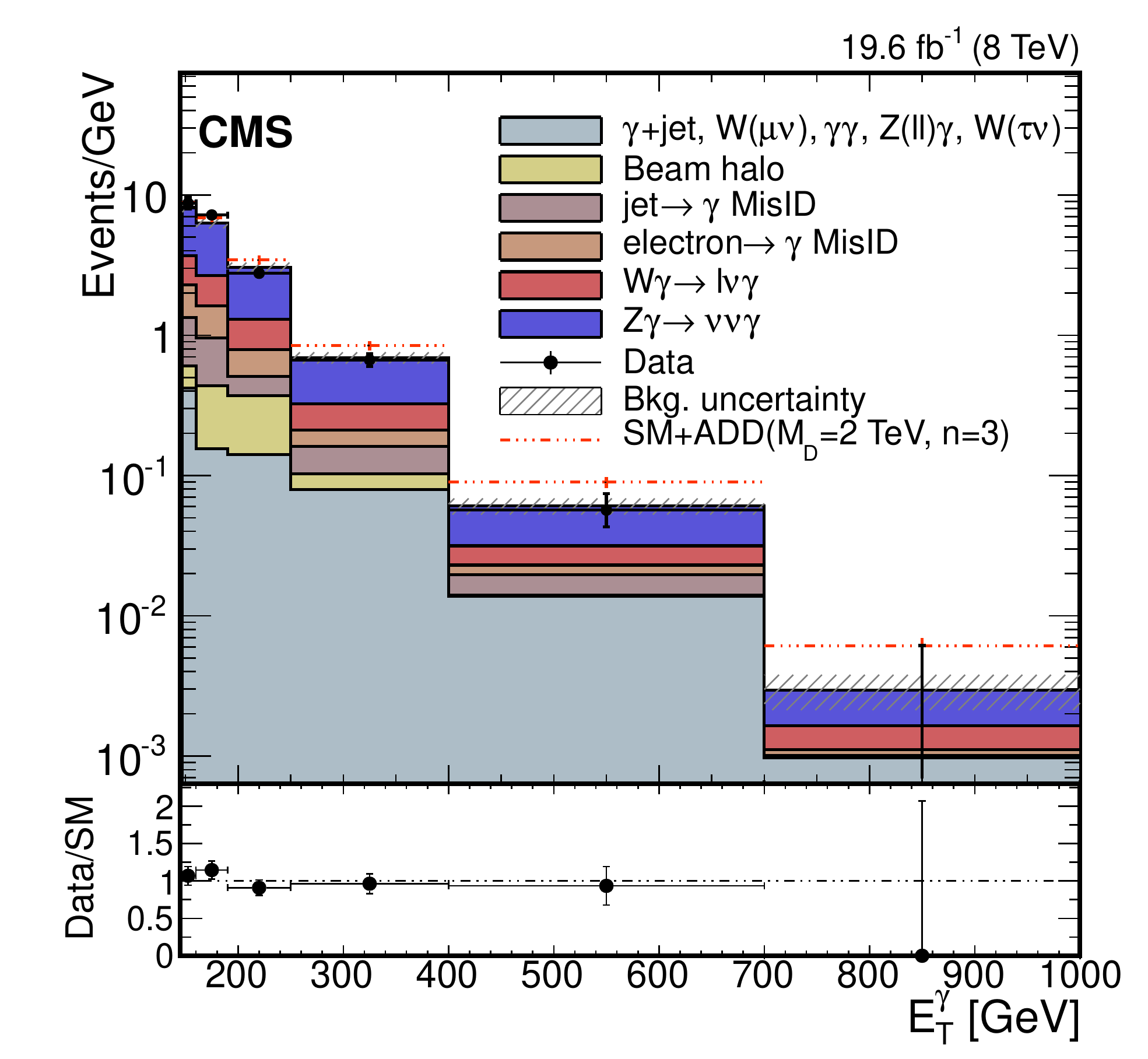}
\includegraphics[width=0.48\textwidth]{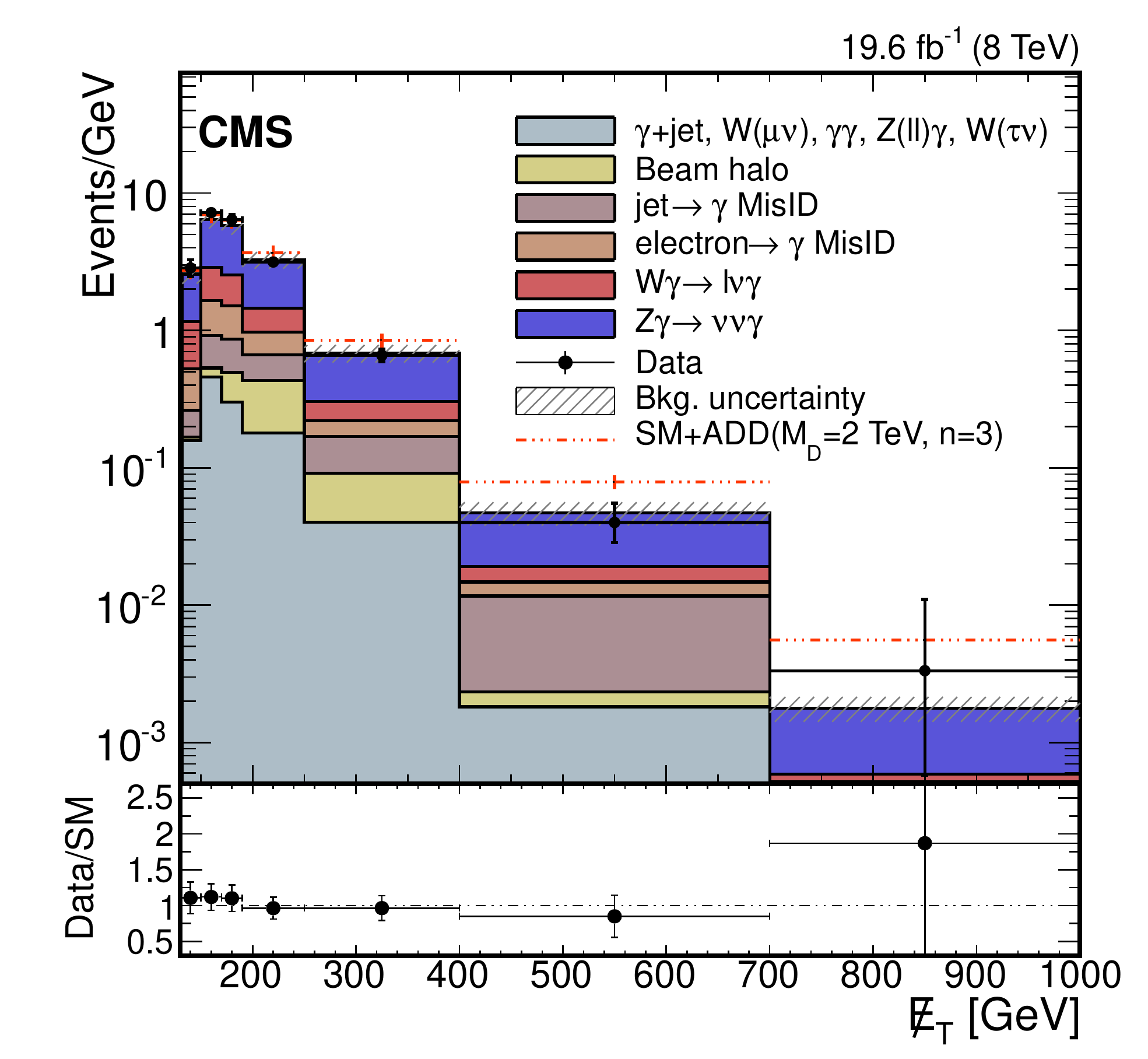}
\caption{The photon $\ET$ and \ETslash\ distributions for the candidate sample, compared with estimated contributions from SM backgrounds, and the predictions from the ADD model
for $\MD=2\TeV$ and $n=3$. The horizontal bar on each data point indicates the width of the bin. The background uncertainty includes statistical and systematic components. The bottom panel shows the ratio of data and SM background predictions.}
\label{fig:stack_plot}
\end{figure}

Table~\ref{tab:BkgSummaryC} shows the estimated number of events and associated uncertainty from each background process along with the total number of events observed in the data, for the entire data set, which corresponds to 19.6\fbinv.
The number of events observed in data agrees with the expectation from SM background.
The photon $\ET$ and \ETslash distributions for the selected candidates
and estimated backgrounds are shown in Fig.~\ref{fig:stack_plot}.
The spectra expected from the ADD model for $\MD = 2\TeV$ and $n = 3$ are also shown for comparison.
Limits are set for the DM, ADD, and branon models using the $\ET^{\gamma}$ spectrum.

\begin{table}[!ht]
\centering
\topcaption{Summary of estimated backgrounds and observed total number of candidates. Backgrounds listed as ``Others'' include the
small contributions from $\PW\to \mu\Pgn$, $\PW\to \tau\Pgn$, $\cPZ\Pgg\to \ell\ell\Pgg$, $\gamma\gamma$, and $\gamma$+jet. Uncertainties
include both statistical and systematic contributions, and the total systematic uncertainty includes the effect of correlations in the individual estimates.
\label{tab:BkgSummaryC}}
{
\begin{tabular}{lc}

Process & Estimate \\
\hline
$\cPZ(\to\nu\bar{\nu})+\gamma    $ & 345 $\pm$ 43  \\
$\PW(\to \ell\nu)+\gamma     $ & 103 $\pm$ 21   \\
$\text{electron}\to\gamma~\mathrm{MisID} $ & 60  $\pm$ 6   \\
$\text{jet}\to\gamma~\mathrm{MisID}$ & 45 $\pm$ 14 \\
Beam halo &  25 $\pm$  6 \\
Others & 36 $\pm$ 3  \\
\hline
Total background                       &  614 $\pm$ 63 \\
\hline
Data                                   & 630    \\

\end{tabular}
}
\end{table}

\begin{figure}[ht!b]
\centering
\includegraphics[width=0.48\textwidth]{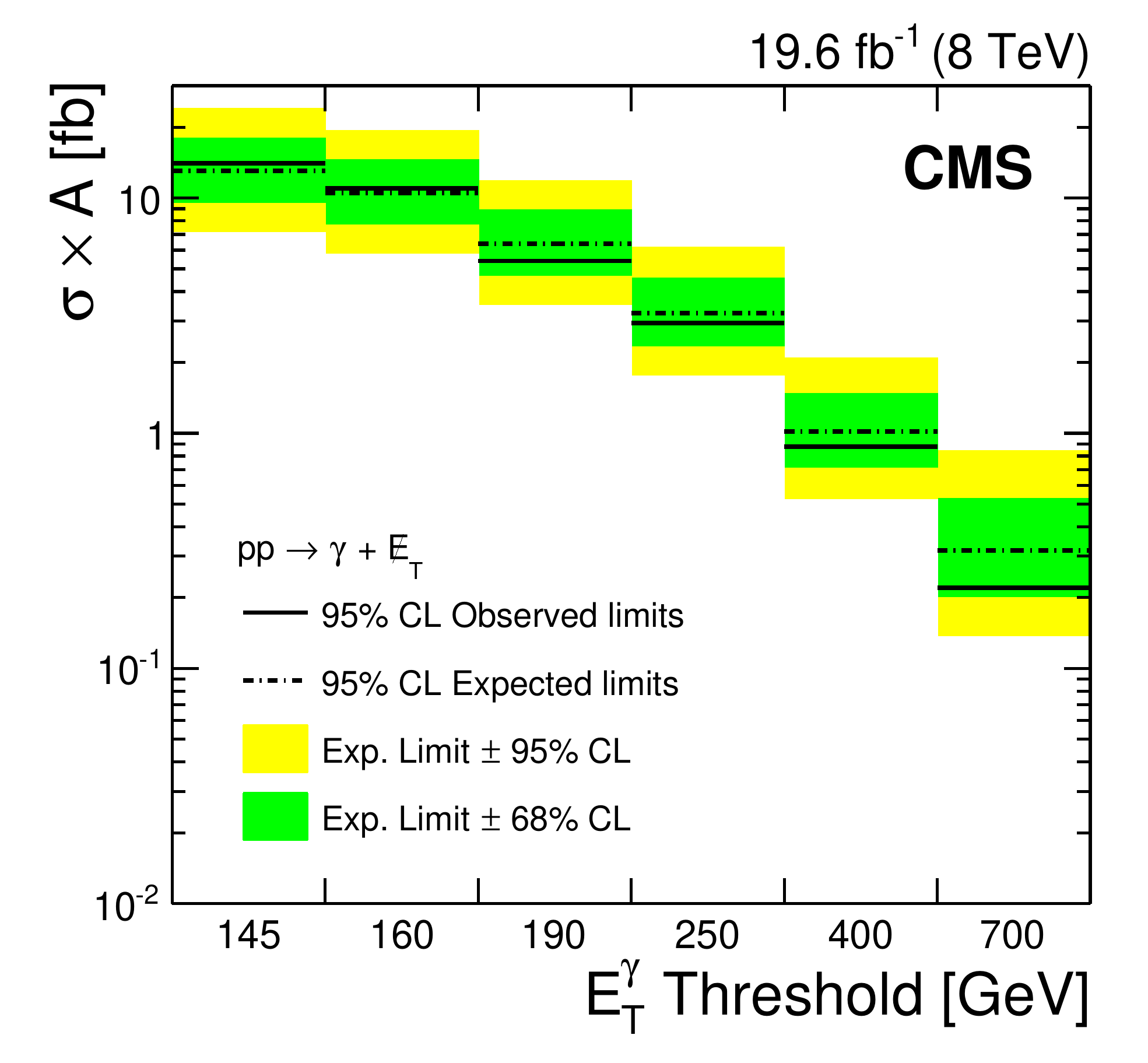}
\caption{Upper limits at 95\% confidence level (CL) on the product of cross section and acceptance as a function of the $\ET^{\gamma}$ threshold ($>$145\GeV) for the photon and \ETslash final state.}\label{fig:MI}

\end{figure}

\begin{table}[htb]
\centering
\topcaption{Observed (expected) 95\% CL and 90\% CL upper limits on  $\sigma  A$ as a function of the cut on the $\ET^{\gamma}$
for the photon and \ETslash final state. The \ETslash threshold is fixed at 140\GeV. In addition to 95\% CL upper limits, 90\% limits are also shown to allow direct comparison
with results from astrophysics DM searches.}
\begin{tabular}{cccc}
$\ET^{\gamma}$ Threshold\,[\GeVns{}] &  $\sigma A$\,[fb] & $\sigma A $\,[fb]  \\
 & (95\% CL)& (90\% CL)\\
\hline
145 & 14 (13)  & 12 (11)\\
160 & 11 (10) &9.3 (8.8) \\
190 & 5.4 (6.4) & 4.4 (5.4) \\
250 & 2.9 (3.2) & 2.4 (2.7) \\
400 & 0.87 (1.0) &0.71 (0.83) \\
700 & 0.22 (0.32) &0.16 (0.25) \\

\end{tabular}
\label{tab:mi}
\end{table}

The product of the acceptance and the efficiency ($A \epsilon$) is estimated by calculating \AeMC from the simulation, and multiplying it by the $F$ to account for the difference in
efficiency between simulation and data. The ADD, DM, and branon simulated samples are processed through the full \GEANTfour-based simulation of the CMS detector \cite{GEANT, GEANTdev}, trigger emulation, and the same event reconstruction programs as used for data.
For DM production, the simulated samples are produced using \MADGRAPH5v1.3.12~\cite{Madgraph}, and requiring $\ET^{\gamma}>130\GeV$ and
$\abs{\eta^{\Pgg}} <1.5$.
The estimated value of \AeMC for $M_{\chi}$ in the range 1--1000\GeV varies over the range 41.6--44.4\% for vector and 41.4--44.1\% for
axial-vector couplings, respectively.
The $\ET^\gamma$ spectra for ADD simulated events are generated using \PYTHIA8.153~\cite{Pythia8}, requiring $\ET^{\gamma}>$ 130\GeV. The \AeMC for the ADD model varies over the range 33.4--37.4\% in the parameter space spanned by $n = 3$--6 and $\MD =1$--3\TeV.
The spectra for simulated branon events are generated using \MADGRAPH5v1.5.5~\cite{Madgraph}, requiring $\ET^{\gamma}>$ 130\GeV.
The value of \AeMC for branon production varies over the range 41.3--48.9\% in the parameter space spanned by the range of branon masses $M_\mathrm{B} =100$--3500\GeV and
brane tensions $f = 100$--1000\GeV. The systematic uncertainty in \AeMC from the modeling of
pileup, the energy calibration, and the resolution for photons, jets,
and \ETslash is $\pm$2.1\%. The systematic uncertainty from the scale factor is 6.4\%, resulting in a total systematic uncertainty in \AeMC of 6.7\%.
The systematic uncertainty in the measured integrated luminosity is $\pm$2.6\%~\cite{CMS-PAS-LUM-13-001}. Theoretical uncertainties in the acceptance of the signal processes, based on the choice of PDF and scale, are found to be of order 1\%, and thus have a negligible effect on the observed limits.

Upper limits on the signal cross section are calculated using the CL$_{\mathrm{s}}$ method~\cite{cls,cls1}. In the fit to the observed spectra, systematic uncertainties are represented by nuisance parameters with log-normal prior probability density functions. The changes in shape of the expected spectra that result from varying the photon energy scale and the theoretical differential cross section within their respective uncertainties are treated using a morphing technique~\cite{vmor}.
The signal region studied in this analysis is defined with the requirement $\ET^{\gamma} >$ 145\GeV.
The observed and expected upper limits on the product of cross section and acceptance ($\sigma  A$), plotted as a function of the $\ET^{\gamma}$ threshold ($>$ 145\GeV), are shown in Fig.~\ref{fig:MI} and listed in Table \ref{tab:mi}. Results shown can be generally applied to any new physics that leads to the photon and \ETslash final state.

\begin{table*}[ht]
\centering
\topcaption{Dark Matter production cross sections as a function of the DM mass, assuming a vector interaction: theoretical DM production cross sections, where the generated photon transverse momentum is greater than 130\GeV and the contact interaction scale $\Lambda$ is 10\TeV; observed (expected) 90\% CL upper limits on the DM production cross section $\sigma$;
90\% CL lower limits
on the contact interaction scale $\Lambda$; and 90\% CL upper limits on the $\chi$-nucleon cross section.}
\begin{tabular}{ccccc}

Mass\,[\GeVns{}] & $\sigma_\mathrm{theo}$\,[fb] & $\sigma$\,[fb] & $\Lambda$\,[\GeVns{}] & $\sigma_{\chi-\text{nucleon}}\,[\unit{cm}^2]$\\
\hline
1  & $ 2.5\times 10^{-4}$ & 7.8 (10.6)  & 750 (694) & $8.2\times 10^{-40}$ ($1.1\times 10^{-39}$) \\
10  & $2.5 \times 10^{-4}$ & 8.0 (10.5)  & 745 (696) & $2.6\times 10^{-39}$ ($3.5\times 10^{-39}$) \\
100  & $2.4\times 10^{-4}$ & 8.0 (11.2)  & 742 (684) & $3.2\times 10^{-39}$ ($4.4\times 10^{-39}$) \\
200  & $2.2\times 10^{-4}$ & 7.6 (9.9)  & 729 (684) & $3.4\times 10^{-39}$ ($4.4\times 10^{-39}$) \\
300  & $1.8\times 10^{-4}$ & 6.9 (9.4)  & 714 (660) & $3.7\times 10^{-39}$ ($5.1\times 10^{-39}$) \\
500  & $1.0\times 10^{-4}$ & 5.2 (7.8)  & 666 (602) & $4.9\times 10^{-39}$ ($7.4\times 10^{-39}$) \\
1000  & $1.5\times 10^{-5}$ & 4.9 (7.2)  & 422 (382) & $3.1\times 10^{-38}$ ($4.6\times 10^{-38}$) \\
\end{tabular}
\label{tab:LimDMV}
\end{table*}

\begin{table*}[ht]
\centering
\topcaption{Dark Matter production cross sections as a function of the DM mass, assuming an axial-vector interaction: theoretical DM production cross sections, where the generated photon transverse momentum is greater than 130\GeV and
 the contact interaction scale $\Lambda$ is 10\TeV; observed (expected) 90\% CL upper limits on the DM production cross section $\sigma$;
90\% CL lower limits
on the contact interaction scale $\Lambda$; and 90\% CL upper limits on the $\chi$-nucleon cross section.}

\begin{tabular}{ccccc}

Mass\,[\GeVns{}] & $\sigma_\text{theo}$\,[fb] & $\sigma$\,[fb] & $\Lambda$\,[\GeVns{}] & $\sigma_{\chi-\text{nucleon}}$\,[cm${^2}$]\\
\hline

1 & $2.4\times 10^{-4}$ & 7.9 (10.5)  & 746 (694) & $3.1\times 10^{-41}$ ($4.1\times 10^{-41}$) \\
10 & $2.5\times 10^{-4}$ & 7.9 (11.0)  & 748 (688) & $9.6\times 10^{-41}$ ($1.3\times 10^{-40}$) \\
100& $2.2\times 10^{-4}$   & 8.2 (10.7)  & 718 (671) & $1.3\times 10^{-40}$ ($1.7\times 10^{-40}$) \\
200 & $1.6\times 10^{-4}$ & 6.7 (9.5)  & 702 (643) & $1.5\times 10^{-40}$ ($2.0\times 10^{-40}$) \\
300 & $1.1\times 10^{-4}$ & 5.8 (8.5)  & 663 (604) & $1.8\times 10^{-40}$ ($2.6\times 10^{-40}$) \\
500 & $4.9\times 10^{-5}$  & 5.5 (8.1)  & 544 (495)&  $4.0\times 10^{-40}$ ($5.9\times 10^{-40}$) \\
1000 & $4.2\times 10^{-6}$  & 5.3 (7.7)  & 298 (272) & $4.5\times 10^{-39}$ ($6.5\times 10^{-39}$) \\

\end{tabular}
\label{tab:LimDMAV}
\end{table*}

\begin{figure}[ht!]
\centering
\includegraphics[width=0.48\textwidth]{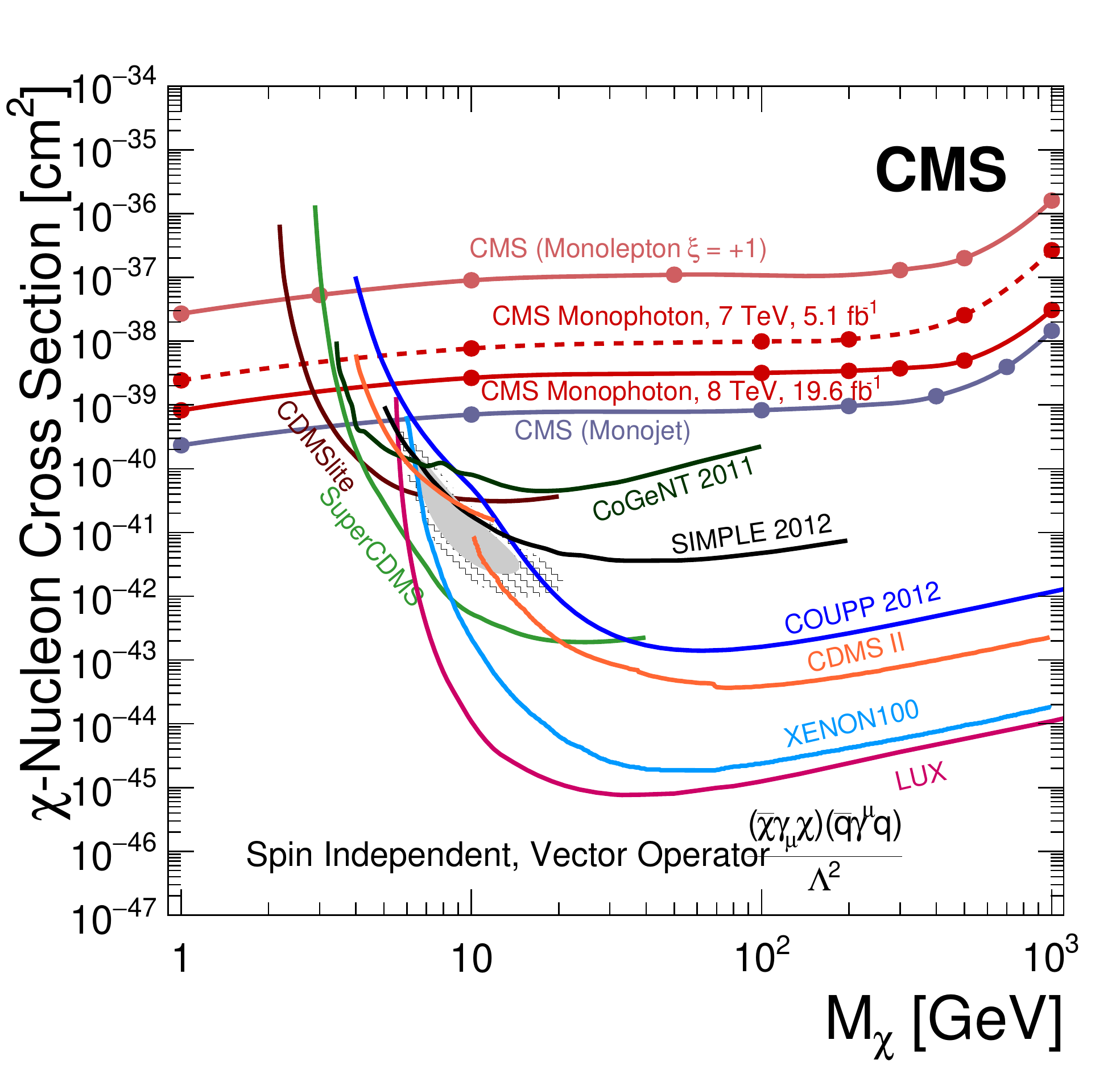}
\includegraphics[width=0.48\textwidth]{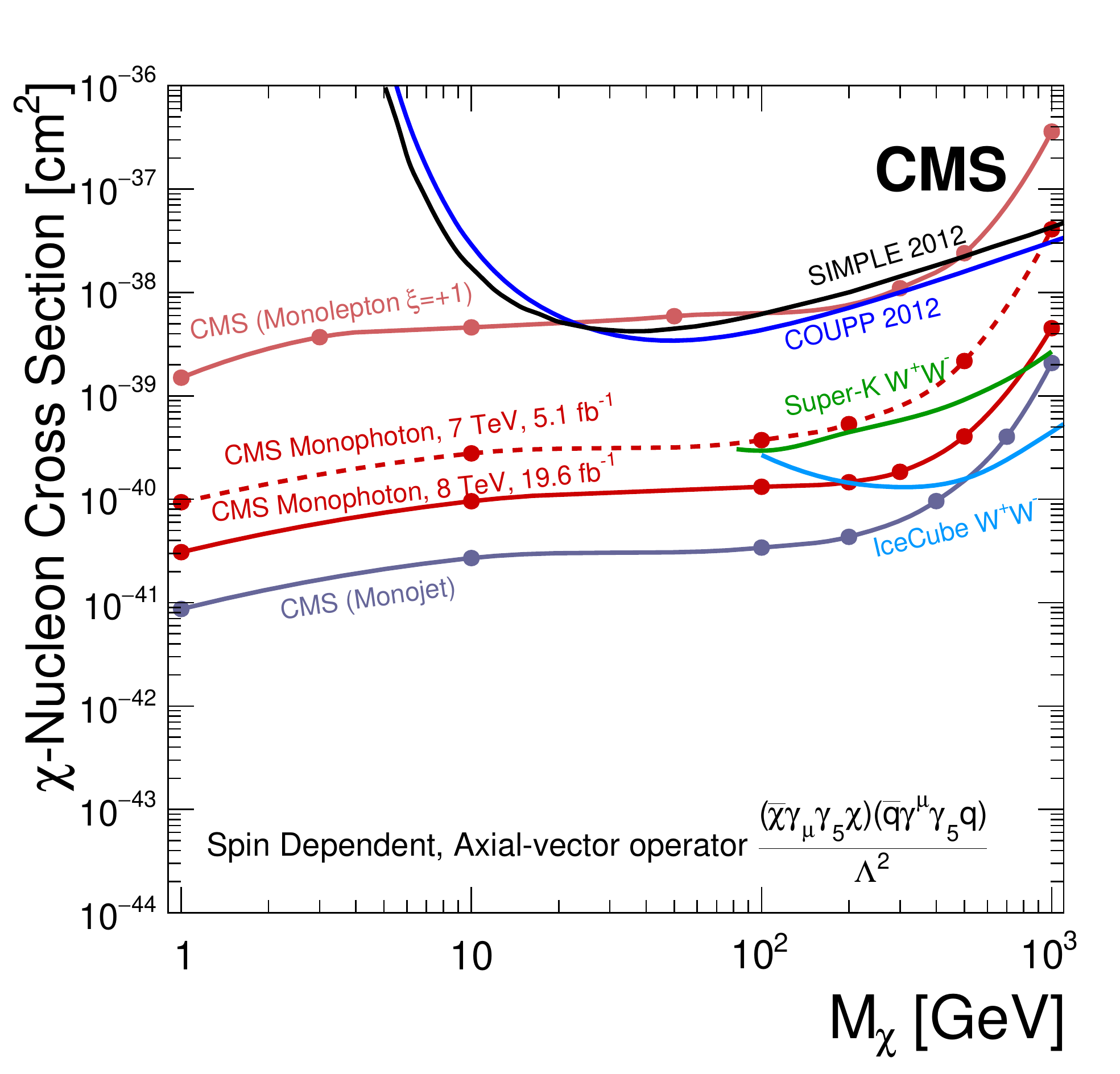}
\caption{The 90\% CL upper limits on the $\chi$-nucleon cross section as a function of the DM particle mass $M_{\chi}$ for spin-independent couplings (\cmsLeft) and spin-dependent couplings (\cmsRight). Results from the current search are shown as ``CMS Monophoton, 8\TeV". Shown are the limits from CMS using monojet~\cite{monojet2014} and monolepton~\cite{monolep} signatures (where $\xi$ is the interference parameter addressing potentially different couplings to up- and down-type quarks and values of $\xi = \pm 1$ maximize the effects of interference). Also shown are the limits from several published direct detection experiments~\cite{XENON100_1, CDMS1, CDMS2, COGENT, SIMPLE, COUPP_1, IceCube_1, SUPERK, LUX, CDMSLITE}. The solid and hatched contours show the 68\% and 95\% CL contours respectively for a possible signal from CDMS~\cite{Agnese:2013rvf}. Limits similar to those from the current search are obtained by ATLAS~\cite{atlas2015}.}
\label{fig:DMLimits}

\end{figure}

\begin{figure}[ht!]
\centering
\includegraphics[width=0.48\textwidth]{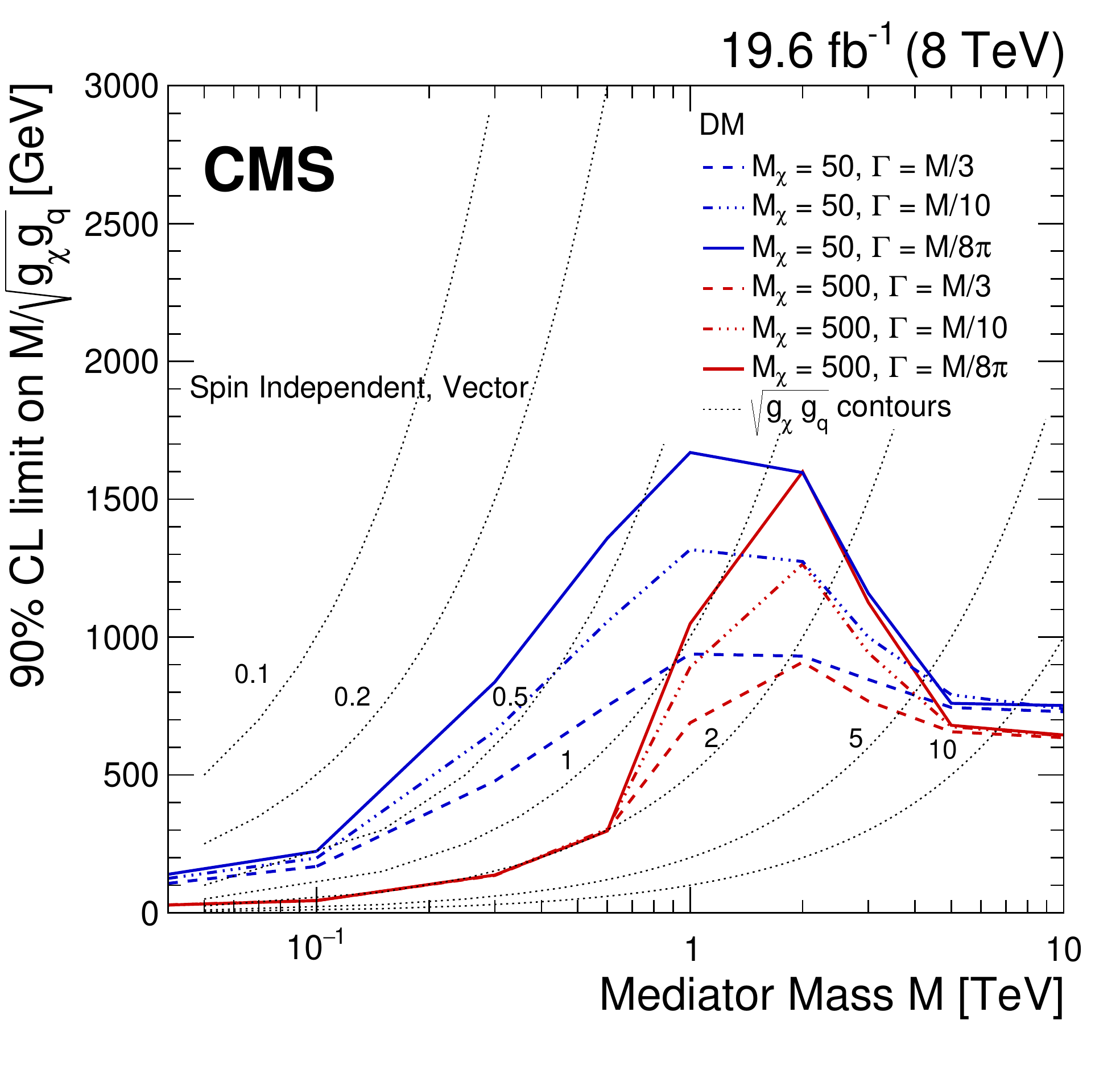}
\caption{Observed limits on the SM-DM interaction mediator mass divided by coupling, $M/\sqrt{g_{\chi} g_{q}}$, as a function of the mediator mass $M$, assuming vector interactions, for DM particle masses of 50\GeV and 500\GeV. The width, $\Gamma$, of the mediator is varied between $M/8\pi$ and $M/3$. The dotted lines show contours of constant coupling.}\label{fig:med}

\end{figure}

Tables~\ref{tab:LimDMV} and \ref{tab:LimDMAV} summarize the 90\% CL upper limits on the production cross sections of the DM particles $\chi\bar{\chi}$,
 as a function of $M_{\chi}$.
In general, the effective operator could be a mixture of vector and axial terms; for explicitness, the limiting cases of pure vector and pure axial vector operators have been chosen, corresponding to spin-independent and spin-dependent interactions, respectively.
Following the procedures of Refs.~\cite{DMTeva} and~\cite{DMCollid}, the upper limits on the DM production cross sections are converted into corresponding lower limits on the contact interaction scale $\Lambda$, which are then translated into upper limits on the $\chi$-nucleon scattering cross sections, calculated within the EFT framework. These results, as a function of $M_{\chi}$, are listed in Tables~\ref{tab:LimDMV} and ~\ref{tab:LimDMAV} and also displayed in Fig.~\ref{fig:DMLimits}.
Superimposed are the results published by other experiments~\cite{XENON100_1, CDMS1, CDMS2, COGENT, SIMPLE, COUPP_1, IceCube_1, SUPERK, LUX, CDMSLITE,Agnese:2013rvf}.

\begin{table}[htbp]
\centering
\topcaption{Observed and expected 95\% CL lower limits on ADD model parameters $\MD$, the effective Planck scale, as a function of $n$, the number of extra dimensions.}
\begin{tabular}{ccc}
$n$ & Obs. Limit [\TeVns{}] & Exp. Limit [\TeVns{}] \\
\hline
3 & 2.12 & 1.96 \\

4 & 2.07& 1.92 \\

5 & 2.02 & 1.89\\

6 & 1.97 & 1.88 \\

\end{tabular}

\label{tab:LimADD}
\end{table}

\begin{figure}[ht!]
\centering
\includegraphics[width=0.48\textwidth]{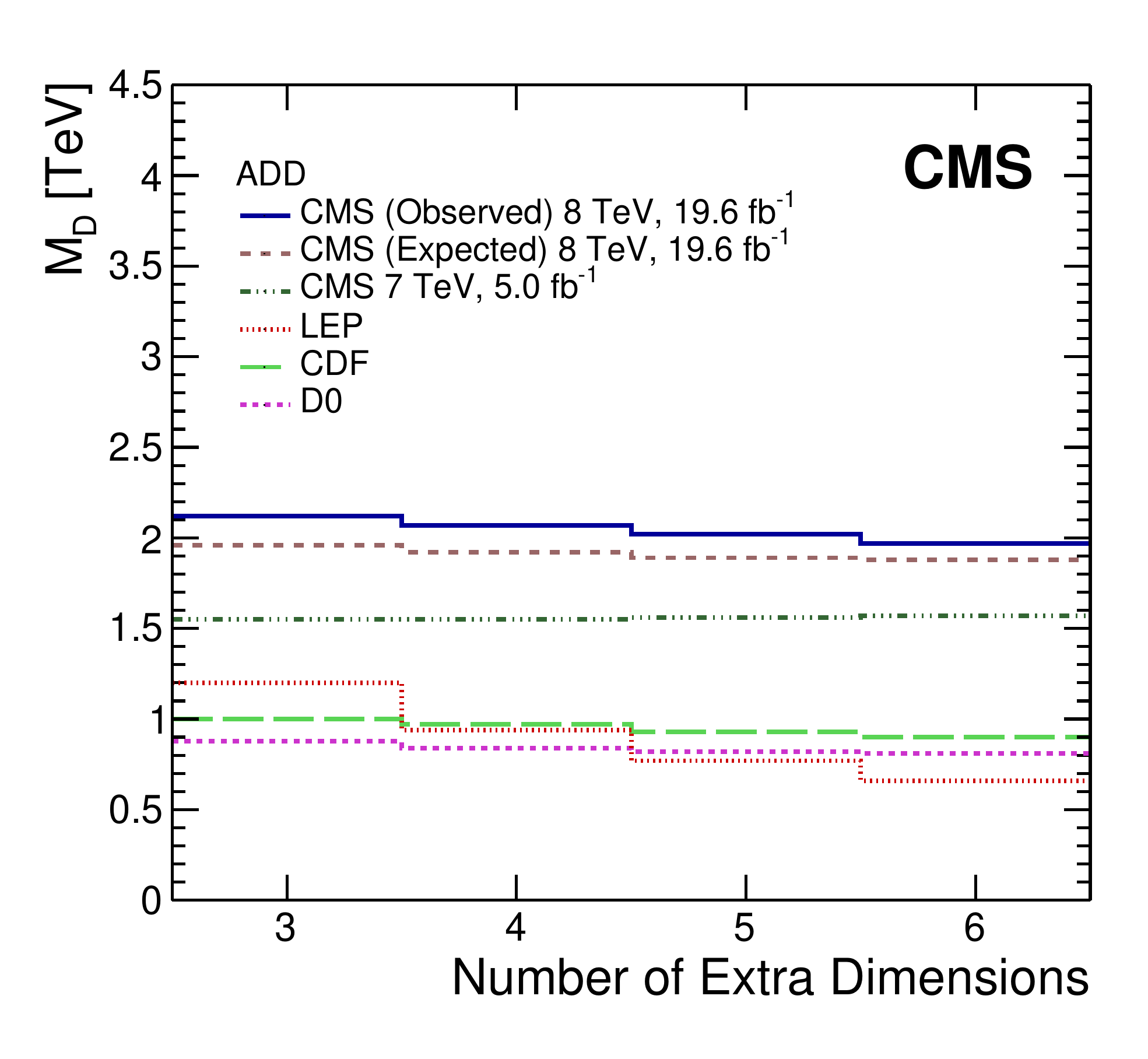}
\caption{The 95\% CL lower limits on the effective Planck scale, $\MD$, as a function of the number of extra dimensions in the ADD model, together with LO results from similar searches at the Tevatron~\cite{CDFgamma, D0gamma}, LEP~\cite{LEPgamma,L3gamma,OPAL,ALEPH} and CMS~\cite{monop}.}\label{fig:ADDLimits}

\end{figure}

The validity of the EFT framework at the energy scale probed by the LHC has been recently
explored in detail~\cite{DMTeva,DMLHC1,DMCollid,An:2012va,Friedland:2011za,Buchmueller:2013dya}.
These studies show that the condition $M \gg Q$ may not always be satisfied because of the high momentum transfer scale at the LHC energies. Therefore, to interpret the data in
a meaningful way where the EFT does not hold, following~\cite{DMLHC1} we consider a simplified model predicting DM
production via an $s$-channel vector mediator. For this simplified model, the simulated samples are produced using \MADGRAPH5v1.5.12~\cite{Madgraph}, and requiring $\ET^{\gamma}>130\GeV$ and $\abs{\eta^{\Pgg}} <1.5$.
Limits on the SM-DM interaction mediator mass divided by coupling, for this model, are shown in Fig.~\ref{fig:med}. The mass of the mediator is varied for two fixed values
of the mass of the DM particle: 50\GeV  and 500\GeV, and the width of the
mediator is varied from $M/8\pi$ to $M/3$~\cite{DMLHC1}. The contours
for fixed values of $\sqrt{g_{\chi}g_\cPq}$ are also shown for comparison. For $M_{\chi} = 500$\GeV the results for a mediator with a mass $\gtrsim$ 5\TeV are similar
to those obtained from the EFT approach as listed in Table~\ref{tab:LimDMV}, while the limits are weaker for $M \lesssim 100$\GeV. The limits are stronger than those of the EFT approach in the range of $M$ from $\sim$100\GeV to $\sim$4\TeV, because of the resonance production enhancement in the cross section. In other words, the limits derived within the EFT framework are conservative in this region. For illustration purposes, similar distributions for $M_{\chi} = 50$\GeV are also shown in Fig.~\ref{fig:med}.

\begin{table*}[h!]
\centering
\topcaption{Observed and expected 95\% CL lower limits on the brane tension $f$ as a function of the branon mass $M_\mathrm{B}$ for $N$=1.}
\begin{tabular}{l|rrrrrrrrrr}
 \multicolumn{1}{c}{}&\multicolumn{10}{c}{$M_\mathrm{B}$ [\GeVns{}]}\\
 \cline{2-11}
  \multicolumn{1}{c}{} & 100& 500& 1000& 1500& 2000& 2500& 2800& 3000& 3200& 3500\\
 \hline
Obs. limit [\GeVns{}] & 410& 380& 320& 240& 170& 97& 59& 48& 36& 20\\
Exp. limit [\GeVns{}] & 400& 370& 310& 240 & 170& 97& 59& 48& 36& 20\\

\end{tabular}

\label{tab:LimBranon}

\end{table*}

\begin{figure}[ht!]
\centering
\includegraphics[width=0.48\textwidth]{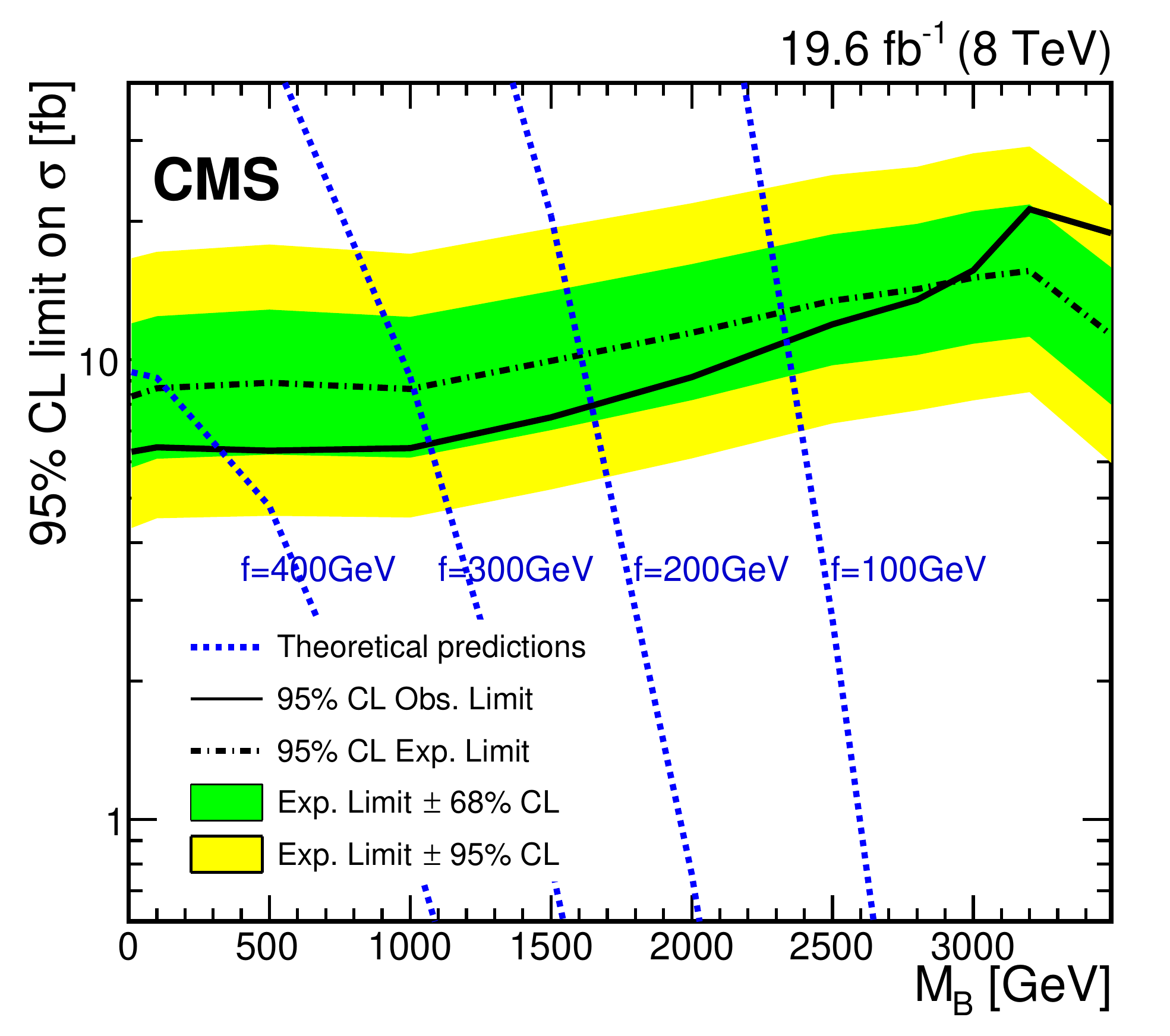}
\includegraphics[width=0.48\textwidth]{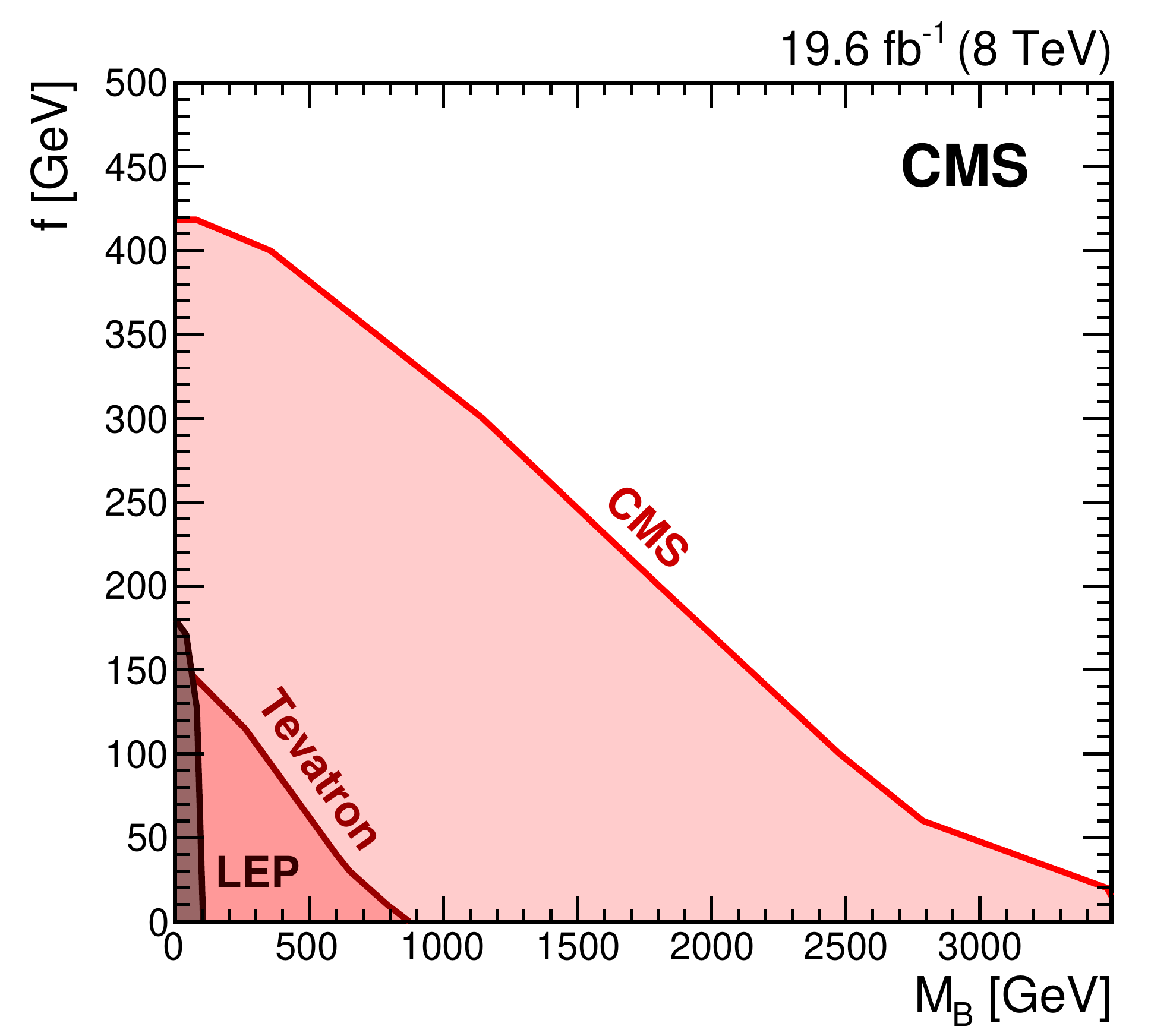}
\caption{The 95\% CL upper limits on the branon cross sections as a function of the branon mass $M_\mathrm{B}$ for $N$=1. Also shown are the theoretical cross sections in the branon model for the brane tension scale $f=100$, 200, 300, and 400\GeV (top).
Limits on $f$ as a function of $M_\mathrm{B}$, compared to results from similar searches at LEP~\cite{l3} and the Tevatron~\cite{cembranos2} (bottom).}
\label{fig:BranonLimits}

\end{figure}

Upper limits at 95\% CL are also placed on the production cross section of the ADD and branon models, and translated into exclusions on the
 parameter space of the models.
For the ADD model we follow the convention of Ref.~\cite{eft_add} and only consider $\hat{s} < M_{D}^{2}$ when calculating the cross sections. The limits on $\MD$ for several values of $n$, the number of extra dimensions, are summarized in Table~\ref{tab:LimADD}.
These limits, along with existing ADD limits from the Tevatron~\cite{D0gamma, CDFgamma} and LEP~\cite{LEPgamma,L3gamma,OPAL,ALEPH}, are shown in
 Fig.~\ref{fig:ADDLimits} as a function of $\MD$. All these results are based on LO cross sections.
Our results extend significantly the experimental limits on the ADD model in the single-photon channel~\cite{monop,Aad:2012fw},
 and set limits of $\MD > 2.12$--1.97\TeV for $n=3$--6, at 95\% CL. These results are comparable with the recent ATLAS limits~\cite{atlas2015}.

Limits on $f$ for branons are summarized in Table~\ref{tab:LimBranon}.
For massless branons, the brane tension $f$ is found to be greater than 410\GeV at 95\% CL.
These limits along with the existing limits from LEP~\cite{l3} and the Tevatron~\cite{cembranos2}, are shown in  Fig.~\ref{fig:BranonLimits}. Branon masses $M_\mathrm{B} < 3.5\TeV$
are excluded at 95\% CL for low brane tension (20\GeV).
These bounds are the most stringent published to date.
These limits complement astrophysical constraints already set on the branon parameters~\cite{branon_dm}.

\section{Summary}
Proton-proton collision events containing a photon and missing transverse momentum have been investigated to search for new phenomena.
In the $\sqrt{s} = 8$\TeV data set corresponding to 19.6\fbinv of integrated luminosity, no deviations from the standard model predictions are observed. Bounds are placed on models predicting monophoton events; specifically, 95\% confidence level upper limits for the cross section times acceptance for the selected final state are set and vary from 14.0\unit{fb} for $\ET^{\gamma}>$ 145\GeV to 0.22\unit{fb} for $\ET^{\gamma}>$ 700\GeV. Constraints are set on $\chi$ production and translated into upper limits on vector and axial-vector contributions to the $\chi$-nucleon scattering cross section, assuming the validity of the EFT framework. For $M_{\chi}$ = 10\GeV, the $\chi$-nucleon cross section is constrained to be less than $2.6\times 10^{-39}\unit{cm}^{2}$ ($9.6\times 10^{-41}\unit{cm}^{2}$) for a spin-independent (spin-dependent) interaction at 90\% confidence level. In addition the most stringent limits to date are obtained on the effective Planck scale in the ADD model with large spatial extra dimensions and on the brane tension scale in the branon model.

\begin{acknowledgments}
We congratulate our colleagues in the CERN accelerator departments for the excellent performance of the LHC and thank the technical and administrative staffs at CERN and at other CMS institutes for their contributions to the success of the CMS effort. In addition, we gratefully acknowledge the computing centers and personnel of the Worldwide LHC Computing Grid for delivering so effectively the computing infrastructure essential to our analyses. Finally, we acknowledge the enduring support for the construction and operation of the LHC and the CMS detector provided by the following funding agencies: BMWFW and FWF (Austria); FNRS and FWO (Belgium); CNPq, CAPES, FAPERJ, and FAPESP (Brazil); MES (Bulgaria); CERN; CAS, MoST, and NSFC (China); COLCIENCIAS (Colombia); MSES and CSF (Croatia); RPF (Cyprus); MoER, ERC IUT and ERDF (Estonia); Academy of Finland, MEC, and HIP (Finland); CEA and CNRS/IN2P3 (France); BMBF, DFG, and HGF (Germany); GSRT (Greece); OTKA and NIH (Hungary); DAE and DST (India); IPM (Iran); SFI (Ireland); INFN (Italy); NRF and WCU (Republic of Korea); LAS (Lithuania); MOE and UM (Malaysia); CINVESTAV, CONACYT, SEP, and UASLP-FAI (Mexico); MBIE (New Zealand); PAEC (Pakistan); MSHE and NSC (Poland); FCT (Portugal); JINR (Dubna); MON, RosAtom, RAS and RFBR (Russia); MESTD (Serbia); SEIDI and CPAN (Spain); Swiss Funding Agencies (Switzerland); MST (Taipei); ThEPCenter, IPST, STAR and NSTDA (Thailand); TUBITAK and TAEK (Turkey); NASU and SFFR (Ukraine); STFC (United Kingdom); DOE and NSF (USA).

Individuals have received support from the Marie-Curie programme and the European Research Council and EPLANET (European Union); the Leventis Foundation; the A. P. Sloan Foundation; the Alexander von Humboldt Foundation; the Belgian Federal Science Policy Office; the Fonds pour la Formation \`a la Recherche dans l'Industrie et dans l'Agriculture (FRIA-Belgium); the Agentschap voor Innovatie door Wetenschap en Technologie (IWT-Belgium); the Ministry of Education, Youth and Sports (MEYS) of the Czech Republic; the Council of Science and Industrial Research, India; the HOMING PLUS programme of Foundation for Polish Science, cofinanced from European Union, Regional Development Fund; the Compagnia di San Paolo (Torino); the Consorzio per la Fisica (Trieste); MIUR project 20108T4XTM (Italy); the Thalis and Aristeia programmes cofinanced by EU-ESF and the Greek NSRF; and the National Priorities Research Program by Qatar National Research Fund.
\end{acknowledgments}
\bibliography{auto_generated}

\cleardoublepage \appendix\section{The CMS Collaboration \label{app:collab}}\begin{sloppypar}\hyphenpenalty=5000\widowpenalty=500\clubpenalty=5000\input{EXO-12-047-authorlist.tex}\end{sloppypar}
\end{document}

%% file: EXO-12-047-authorlist.tex
\textbf{Yerevan Physics Institute,  Yerevan,  Armenia}\\*[0pt]
V.~Khachatryan, A.M.~Sirunyan, A.~Tumasyan
\vskip\cmsinstskip
\textbf{Institut f\"{u}r Hochenergiephysik der OeAW,  Wien,  Austria}\\*[0pt]
W.~Adam, T.~Bergauer, M.~Dragicevic, J.~Er\"{o}, M.~Friedl, R.~Fr\"{u}hwirth\cmsAuthorMark{1}, V.M.~Ghete, C.~Hartl, N.~H\"{o}rmann, J.~Hrubec, M.~Jeitler\cmsAuthorMark{1}, W.~Kiesenhofer, V.~Kn\"{u}nz, M.~Krammer\cmsAuthorMark{1}, I.~Kr\"{a}tschmer, D.~Liko, I.~Mikulec, D.~Rabady\cmsAuthorMark{2}, B.~Rahbaran, H.~Rohringer, R.~Sch\"{o}fbeck, J.~Strauss, W.~Treberer-Treberspurg, W.~Waltenberger, C.-E.~Wulz\cmsAuthorMark{1}
% ...
\vskip\cmsinstskip
% ...
1:~~Also at Vienna University of Technology, Vienna, Austria\\
2:~~Also at CERN, European Organization for Nuclear Research, Geneva, Switzerland\\
% ...


%% file: EXO-12-047_temp.bbl
\providecommand{\href}[2]{#2}\begingroup\raggedright\begin{thebibliography}{10}%
\makeatletter
\providecommand{\hrefCMSnoop }[0]{\@secondoftwo}%
\makeatother
\providecommand{\doi}{\texttt{doi:}\begingroup \urlstyle{tt}\Url}

\bibitem{DMGeneral}
\hrefCMSnoop {}{R.~Gaitskell, ``Direct Detection of Dark Matter'',} \textit{
  Annual Review of Nuclear and Particle Science} \textbf{ 54} (2004) 315,
  \href{http://dx.doi.org/10.1146/annurev.nucl.54.070103.181244}{\doi{10.1146/annurev.nucl.54.070103.181244}}.

\bibitem{DMTeva}
\hrefCMSnoop {}{Y.~Bai, P.~J. Fox, and R.~Harnik, ``The {Tevatron} at the
  frontier of dark matter direct detection'',} \textit{ JHEP} \textbf{ 12}
  (2010) 048,
  \href{http://dx.doi.org/10.1007/JHEP12(2010)048}{\doi{10.1007/JHEP12(2010)048}},
  \href{http://www.arXiv.org/abs/1005.3797v2}{\texttt{arXiv:1005.3797v2}}.

\bibitem{DMLHC1}
\hrefCMSnoop {}{P.~J. Fox, R.~Harnik, J.~Kopp, and Y.~Tsai, ``{Missing energy
  signatures of dark matter at the LHC}'',} \textit{ Phys. Rev. D} \textbf{ 85}
  (2012) 056011,
  \href{http://dx.doi.org/10.1103/PhysRevD.85.056011}{\doi{10.1103/PhysRevD.85.056011}},
\href{http://www.arXiv.org/abs/1109.4398}{\texttt{arXiv:1109.4398}}.

\bibitem{Majorana_dm}
J.~Goodman\hrefCMSnoop {}{ {et~al.}, ``Constraints on light {M}ajorana dark
  matter from colliders'',} \textit{ Phys. Lett. B} \textbf{ 695} (2011) 185,
  \href{http://dx.doi.org/10.1016/j.physletb.2010.11.009}{\doi{10.1016/j.physletb.2010.11.009}},
\href{http://www.arXiv.org/abs/1005.1286}{\texttt{arXiv:1005.1286}}.

\bibitem{DMCollid}
J.~Goodman\hrefCMSnoop {}{ {et~al.}, ``Constraints on dark matter from
  colliders'',} \textit{ Phys. Rev. D} \textbf{ 82} (2010) 116010,
  \href{http://dx.doi.org/10.1103/PhysRevD.82.116010}{\doi{10.1103/PhysRevD.82.116010}},
\href{http://www.arXiv.org/abs/1008.1783}{\texttt{arXiv:1008.1783}}.

\bibitem{ADD}
\hrefCMSnoop {}{N.~Arkani-Hamed, S.~Dimopoulos, and G.~R. Dvali, ``{The
  hierarchy problem and new dimensions at a millimeter}'',} \textit{ Phys.
  Lett. B} \textbf{ 429} (1998) 263,
  \href{http://dx.doi.org/10.1016/S0370-2693(98)00466-3}{\doi{10.1016/S0370-2693(98)00466-3}},
  \href{http://www.arXiv.org/abs/hep-ph/9803315}{\texttt{arXiv:hep-ph/9803315}}.

\bibitem{ADD1}
\hrefCMSnoop {}{N.~Arkani-Hamed, S.~Dimopoulos, and G.~R. Dvali,
  ``{Phenomenology, astrophysics and cosmology of theories with submillimeter
  dimensions and TeV scale quantum gravity}'',} \textit{ Phys. Rev. D} \textbf{
  59} (1999) 086004,
  \href{http://dx.doi.org/10.1103/PhysRevD.59.086004}{\doi{10.1103/PhysRevD.59.086004}},
\href{http://www.arXiv.org/abs/hep-ph/9807344}{\texttt{arXiv:hep-ph/9807344}}.

\bibitem{sundrum}
\hrefCMSnoop {}{R.~Sundrum, ``{Effective field theory for a three-brane
  universe}'',} \textit{ Phys. Rev. D} \textbf{ 59} (1999) 085009,
  \href{http://dx.doi.org/10.1103/PhysRevD.59.085009}{\doi{10.1103/PhysRevD.59.085009}},
\href{http://www.arXiv.org/abs/hep-ph/9805471}{\texttt{arXiv:hep-ph/9805471}}.

\bibitem{dobado}
\hrefCMSnoop {}{A.~Dobado and A.~L. Maroto, ``{The dynamics of the Goldstone
  bosons on the brane}'',} \textit{ Nucl. Phys. B} \textbf{ 592} (2001) 203,
  \href{http://dx.doi.org/10.1016/S0550-3213(00)00574-5}{\doi{10.1016/S0550-3213(00)00574-5}},
\href{http://www.arXiv.org/abs/hep-ph/0007100}{\texttt{arXiv:hep-ph/0007100}}.

\bibitem{cembranos1}
\hrefCMSnoop {}{J.~A.~R. Cembranos, A.~Dobado, and A.~L. Maroto, ``Brane
  skyrmions and wrapped states'',} \textit{ Phys. Rev. D} \textbf{ 65} (2002)
  026005,
  \href{http://dx.doi.org/10.1103/PhysRevD.65.026005}{\doi{10.1103/PhysRevD.65.026005}},
\href{http://www.arXiv.org/abs/hep-ph/0106322}{\texttt{arXiv:hep-ph/0106322}}.

\bibitem{cembranos3}
\hrefCMSnoop {}{J.~A.~R. Cembranos, R.~L. Delgado, and A.~Dobado, ``{Brane
  worlds at the LHC: branons and KK gravitons}'',} \textit{ Phys. Rev. D}
  \textbf{ 88} (2013) 075021,
  \href{http://dx.doi.org/10.1103/PhysRevD.88.075021}{\doi{10.1103/PhysRevD.88.075021}},
\href{http://www.arXiv.org/abs/1306.4900}{\texttt{arXiv:1306.4900}}.

\bibitem{branon_dm}
\hrefCMSnoop {}{J.~A.~R. Cembranos, A.~Dobado, and A.~L. Maroto, ``Cosmological
  and astrophysical limits on brane fluctuations'',} \textit{ Phys. Rev. D}
  \textbf{ 68} (2003) 103505,
  \href{http://dx.doi.org/10.1103/PhysRevD.68.103505}{\doi{10.1103/PhysRevD.68.103505}},
\href{http://www.arXiv.org/abs/hep-ph/0307062}{\texttt{arXiv:hep-ph/0307062}}.

\bibitem{cembranos2}
\hrefCMSnoop {}{J.~A.~R. Cembranos, A.~Dobado, and A.~L. Maroto, ``Branon
  search in hadronic colliders'',} \textit{ Phys. Rev. D} \textbf{ 70} (2004)
  096001,
  \href{http://dx.doi.org/10.1103/PhysRevD.70.096001}{\doi{10.1103/PhysRevD.70.096001}},
\href{http://www.arXiv.org/abs/hep-ph/0405286}{\texttt{arXiv:hep-ph/0405286}}.

\bibitem{Chatrchyan:2013dga}
\hrefCMSnoop {}{{CMS Collaboration}, ``{Energy calibration and resolution of
  the CMS electromagnetic calorimeter in pp collisions at $\sqrt{s}$ = 7
  TeV}'',} \textit{ JINST} \textbf{ 8} (2013) P09009,
  \href{http://dx.doi.org/10.1088/1748-0221/8/09/P09009}{\doi{10.1088/1748-0221/8/09/P09009}},
\href{http://www.arXiv.org/abs/1306.2016}{\texttt{arXiv:1306.2016}}.

\bibitem{JINST1}
\hrefCMSnoop {}{{CMS Collaboration}, ``The {CMS} experiment at the {CERN}
  {LHC}'',} \textit{ JINST} \textbf{ 3} (2008) S08004,
  \href{http://dx.doi.org/10.1088/1748-0221/3/08/S08004}{\doi{10.1088/1748-0221/3/08/S08004}}.

\bibitem{purity}
\hrefCMSnoop {}{{CMS Collaboration}, ``{Performance of photon reconstruction
  and identification with the CMS detector in proton-proton collisions at
  $\sqrt{s}=8$ TeV}'',} (2015).
  \href{http://www.arXiv.org/abs/1502.02702}{\texttt{arXiv:1502.02702}}.
Submitted to JINST.

\bibitem{CMS-PAS-EGM-10-006}
\href {http://cdsweb.cern.ch/record/1324545}{{CMS Collaboration}, ``Isolated
  Photon Reconstruction and Identification at $\sqrt{s}=7$~TeV'',} CMS Physics
  Analysis Summary CMS-PAS-EGM-10-006, 2011.

\bibitem{CMS-PAS-PFT-09-001}
\href {http://cdsweb.cern.ch/record/1194487}{{CMS Collaboration},
  ``{Particle-Flow Event Reconstruction in CMS and Performance for Jets, Taus,
  and MET}'',} CMS Physics Analysis Summary CMS-PAS-PFT-09-001, 2009.

\bibitem{CMS-PAS-PFT-10-001}
\href {http://cdsweb.cern.ch/record/1247373}{{CMS Collaboration},
  ``{Commissioning of the Particle-flow Event Reconstruction with the first LHC
  collisions recorded in the CMS detector}'',} CMS Physics Analysis Summary
  CMS-PAS-PFT-10-001, 2010.

\bibitem{FastJet1}
\hrefCMSnoop {}{M.~Cacciari, G.~P. Salam, and G.~Soyez, ``The catchment area of
  jets'',} \textit{ JHEP} \textbf{ 04} (2008) 005,
  \href{http://dx.doi.org/10.1088/1126-6708/2008/04/005}{\doi{10.1088/1126-6708/2008/04/005}},
\href{http://www.arXiv.org/abs/0802.1188}{\texttt{arXiv:0802.1188}}.

\bibitem{Petyt:2012upa}
\hrefCMSnoop {}{{CMS Collaboration}, ``{Mitigation of anomalous APD signals in
  the CMS electromagnetic calorimeter}'',} in \textit{ XVth International
  Conference on Calorimetry in High Energy Physics (CALOR2012)}.
\newblock Santa Fe, USA, June, 2012.
\newblock {[J. Phys. Conf. Ser. 404 (2012) 012043]}.
\href{http://dx.doi.org/10.1088/1742-6596/404/1/012043}{\doi{10.1088/1742-6596/404/1/012043}}.

\bibitem{EGM-10-004}
\href {http://cdsweb.cern.ch/record/1299116}{{CMS Collaboration}, ``Electron
  Reconstruction and Identification at $\sqrt{s}$ = 7~TeV'',} CMS Physics
  Analysis Summary CMS-PAS-EGM-10-004, 2010.

\bibitem{Cacciari:2008gp}
\hrefCMSnoop {}{M.~Cacciari, G.~P. Salam, and G.~Soyez, ``{The anti-$k_t$ jet
  clustering algorithm}'',} \textit{ JHEP} \textbf{ 04} (2008) 063,
  \href{http://dx.doi.org/10.1088/1126-6708/2008/04/063}{\doi{10.1088/1126-6708/2008/04/063}},
  \href{http://www.arXiv.org/abs/0802.1189}{\texttt{arXiv:0802.1189}}.

\bibitem{CMS-PAS-JME-13-005}
\href {http://cdsweb.cern.ch/record/1581583}{{CMS Collaboration}, ``Pileup Jet
  Identification'',} CMS Physics Analysis Summary CMS-PAS-JME-13-005, 2013.

\bibitem{metreso}
\href {http://cdsweb.cern.ch/record/1543527}{{CMS Collaboration}, ``{MET
  performance in 8 TeV data}'',} CMS Physics Analysis Summary
  CMS-PAS-JME-12-002, 2013.

\bibitem{GEANT}
\hrefCMSnoop {}{{GEANT4} Collaboration, ``GEANT4---a simulation toolkit'',}
  \textit{ Nucl. Instrum. Meth. A} \textbf{ 506} (2003) 250,
  \href{http://dx.doi.org/10.1016/S0168-9002(03)01368-8}{\doi{10.1016/S0168-9002(03)01368-8}}.

\bibitem{GEANTdev}
\hrefCMSnoop {}{J.~Allison {et~al.}, ``Geant4 developments and applications'',}
  \textit{ IEEE Trans. Nucl. Sci.} \textbf{ 53} (2006) 270,
  \href{http://dx.doi.org/10.1109/TNS.2006.869826}{\doi{10.1109/TNS.2006.869826}}.

\bibitem{Madgraph_new}
J.~Alwall\hrefCMSnoop {}{ {et~al.}, ``The automated computation of tree-level
  and next-to-leading order differential cross sections, and their matching to
  parton shower simulations'',} \textit{ JHEP} \textbf{ 07} (2014) 079,
  \href{http://dx.doi.org/10.1007/JHEP07(2014)079}{\doi{10.1007/JHEP07(2014)079}},
\href{http://www.arXiv.org/abs/1405.0301}{\texttt{arXiv:1405.0301}}.

\bibitem{MCFM}
J.~Campbell, R.~Ellis, and C.~Williams, ``MCFM v6.1: A Monte Carlo for
  FeMtobarn processes at Hadron Colliders'', 2011, \url
  {http://mcfm.fnal.gov/mcfm.pdf}.

\bibitem{Alekhin:2011sk}
\hrefCMSnoop {}{S.~Alekhin {et~al.}, ``{The PDF4LHC Working Group Interim
  Report}'',} (2011).
\href{http://www.arXiv.org/abs/1101.0536}{\texttt{arXiv:1101.0536}}.

\bibitem{PDF4LHC}
M.~Botje\hrefCMSnoop {}{ {et~al.}, ``{The PDF4LHC Working Group Interim
  Recommendations}'',} (2011).
  \href{http://www.arXiv.org/abs/1101.0538}{\texttt{arXiv:1101.0538}}.

\bibitem{PDF4LHC1}
\hrefCMSnoop {}{{NNPDF} Collaboration, ``Impact of heavy quark masses on parton
  distributions and {LHC} phenomenology'',} \textit{ Nucl. Phys. B} \textbf{
  849} (2011) 296,
  \href{http://dx.doi.org/10.1016/j.nuclphysb.2011.03.021}{\doi{10.1016/j.nuclphysb.2011.03.021}},
\href{http://www.arXiv.org/abs/1101.1300}{\texttt{arXiv:1101.1300}}.

\bibitem{Pythia6}
\hrefCMSnoop {}{T.~Sj{\"{o}}strand, S.~Mrenna, and P.~Z. Skands, ``PYTHIA 6.4
  physics and manual'',} \textit{ JHEP} \textbf{ 05} (2006) 26,
  \href{http://dx.doi.org/10.1088/1126-6708/2006/05/026}{\doi{10.1088/1126-6708/2006/05/026}},
  \href{http://www.arXiv.org/abs/hep-ph/0603175}{\texttt{arXiv:hep-ph/0603175}}.

\bibitem{CTEQ6L1}
J.~Pumplin\hrefCMSnoop {}{ {et~al.}, ``{New generation of parton distributions
  with uncertainties from global QCD analysis}'',} \textit{ JHEP} \textbf{ 07}
  (2002) 012,
  \href{http://dx.doi.org/10.1088/1126-6708/2002/07/012}{\doi{10.1088/1126-6708/2002/07/012}},
\href{http://www.arXiv.org/abs/hep-ph/0201195}{\texttt{arXiv:hep-ph/0201195}}.

\bibitem{JetEnCor2011V2}
\hrefCMSnoop {}{{CMS Collaboration}, ``Determination of jet energy calibration
  and transverse momentum resolution in {CMS}'',} \textit{ JINST} \textbf{ 6}
  (2011) P11002,
  \href{http://dx.doi.org/10.1088/1748-0221/6/11/P11002}{\doi{10.1088/1748-0221/6/11/P11002}},
  \href{http://www.arXiv.org/abs/1107.4277}{\texttt{arXiv:1107.4277}}.

\bibitem{JME-10-009}
\hrefCMSnoop {}{{CMS Collaboration}, ``Missing transverse energy performance of
  the {CMS} detector'',} \textit{ J. Instrum.} \textbf{ 6} (2011) P09001,
  \href{http://dx.doi.org/10.1088/1748-0221/6/09/P09001}{\doi{10.1088/1748-0221/6/09/P09001}}.

\bibitem{monojet2014}
\hrefCMSnoop {}{{CMS Collaboration}, ``{Search for dark matter, extra
  dimensions, and unparticles in monojet events in proton-proton collisions at
  $\sqrt{s}$ = 8 TeV}'',} (2014).
  \href{http://www.arXiv.org/abs/1408.3583}{\texttt{arXiv:1408.3583}}.
Submitted to Eur. Phys. J. C.

\bibitem{tp}
\hrefCMSnoop {}{{CMS Collaboration}, ``Measurement of the inclusive {W} and {Z}
  production cross sections in pp collisions at {$\sqrt{s}=7\TeV$} with the CMS
  experiment'',} \textit{ J. High Energy Phys.} \textbf{ 10} (2011) 132,
  \href{http://dx.doi.org/10.1007/JHEP10(2011)132}{\doi{10.1007/JHEP10(2011)132}}.

\bibitem{Madgraph}
J.~Alwall\hrefCMSnoop {}{ {et~al.}, ``{MadGraph 5: going beyond}'',} \textit{
  JHEP} \textbf{ 06} (2011) 128,
  \href{http://dx.doi.org/10.1007/JHEP06(2011)128}{\doi{10.1007/JHEP06(2011)128}},
\href{http://www.arXiv.org/abs/1106.0522}{\texttt{arXiv:1106.0522}}.

\bibitem{Pythia8}
\hrefCMSnoop {}{T.~Sj{\"o}strand, S.~Mrenna, and P.~Skands, ``A brief
  introduction to {PYTHIA} 8.1'',} \textit{ Comput. Phys. Commun.} \textbf{
  178} (2008) 852,
  \href{http://dx.doi.org/10.1016/j.cpc.2008.01.036}{\doi{10.1016/j.cpc.2008.01.036}},
  \href{http://www.arXiv.org/abs/0710.3820}{\texttt{arXiv:0710.3820}}.

\bibitem{CMS-PAS-LUM-13-001}
\href {http://cdsweb.cern.ch/record/1598864}{{CMS Collaboration}, ``CMS
  Luminosity Based on Pixel Cluster Counting - Summer 2013 Update'',} CMS
  Physics Analysis Summary CMS-PAS-LUM-13-001, 2013.

\bibitem{cls}
\hrefCMSnoop {}{A.~L. Read, ``Presentation of search results: the {$CL_s$}
  technique'',} \textit{ J. Phys. G} \textbf{ 28} (2002) 2693,
\href{http://dx.doi.org/10.1088/0954-3899/28/10/313}{\doi{10.1088/0954-3899/28/10/313}}.

\bibitem{cls1}
\hrefCMSnoop {}{T.~Junk, ``{Confidence level computation for combining searches
  with small statistics}'',} \textit{ Nucl. Instrum. Meth. A} \textbf{ 434}
  (1999) 435,
  \href{http://dx.doi.org/10.1016/S0168-9002(99)00498-2}{\doi{10.1016/S0168-9002(99)00498-2}},
\href{http://www.arXiv.org/abs/hep-ex/9902006}{\texttt{arXiv:hep-ex/9902006}}.

\bibitem{vmor}
\hrefCMSnoop {}{J.~S. Conway, ``Nuisance Parameters in Likelihoods for
  Multisource Spectra'',} in \textit{ Proceedings of PHYSTAT 2011 Workshop on
  Statistical Issues Related to Discovery Claims in Search Experiments and
  Unfolding}, H.~B. Propser and L.~Lyons, eds., p.~115.
\newblock CERN, Geneva, Switzerland, January 2011, 2011.
\newblock
  \href{http://dx.doi.org/10.5170/CERN-2011-006}{\doi{10.5170/CERN-2011-006}}.

\bibitem{monolep}
\hrefCMSnoop {}{{CMS Collaboration}, ``{Search for physics beyond the standard
  model in final states with a lepton and missing transverse energy in
  proton-proton collisions at $\sqrt{s} = 8$ TeV}'',} (2014).
  \href{http://www.arXiv.org/abs/1408.2745}{\texttt{arXiv:1408.2745}}.
Submitted to Phys. Rev. D.

\bibitem{XENON100_1}
\hrefCMSnoop {}{{XENON100} Collaboration, ``{Dark Matter Results from 225 Live
  Days of XENON100 Data}'',} \textit{ Phys. Rev. Lett.} \textbf{ 109} (2012)
  181301,
  \href{http://dx.doi.org/10.1103/PhysRevLett.109.181301}{\doi{10.1103/PhysRevLett.109.181301}},
\href{http://www.arXiv.org/abs/1207.5988}{\texttt{arXiv:1207.5988}}.

\bibitem{CDMS1}
\hrefCMSnoop {}{{CDMS} Collaboration, ``Results from a Low-Energy Analysis of
  the {CDMS II} Germanium Data'',} \textit{ Phys. Rev. Lett.} \textbf{ 106}
  (2011) 131302,
  \href{http://dx.doi.org/10.1103/PhysRevLett.106.131302}{\doi{10.1103/PhysRevLett.106.131302}},
  \href{http://www.arXiv.org/abs/1011.2482v3}{\texttt{arXiv:1011.2482v3}}.

\bibitem{CDMS2}
\hrefCMSnoop {}{{CDMS II} Collaboration, ``Dark Matter Search Results from the
  {CDMS II} Experiment'',} \textit{ Science} \textbf{ 327} (2010) 1619,
  \href{http://dx.doi.org/10.1126/science.1186112}{\doi{10.1126/science.1186112}}.

\bibitem{COGENT}
\hrefCMSnoop {}{{CoGeNT} Collaboration, ``Results from a Search for Light-Mass
  Dark Matter with a $p$-Type Point Contact Germanium Detector'',} \textit{
  Phys. Rev. Lett.} \textbf{ 106} (2011) 131301,
  \href{http://dx.doi.org/10.1103/PhysRevLett.106.131301}{\doi{10.1103/PhysRevLett.106.131301}},
  \href{http://www.arXiv.org/abs/1002.4703}{\texttt{arXiv:1002.4703}}.

\bibitem{SIMPLE}
\hrefCMSnoop {}{{SIMPLE} Collaboration, ``{Final Analysis and Results of the
  Phase II SIMPLE Dark Matter Search}'',} \textit{ Phys. Rev. Lett.} \textbf{
  108} (2012) 201302,
  \href{http://dx.doi.org/10.1103/PhysRevLett.108.201302}{\doi{10.1103/PhysRevLett.108.201302}},
\href{http://www.arXiv.org/abs/1106.3014}{\texttt{arXiv:1106.3014}}.

\bibitem{COUPP_1}
\hrefCMSnoop {}{{COUPP} Collaboration, ``{First Dark Matter Search Results from
  a 4-kg CF$_3$I Bubble Chamber Operated in a Deep Underground Site}'',}
  \textit{ Phys. Rev. D} \textbf{ 86} (2012) 052001,
  \href{http://dx.doi.org/10.1103/PhysRevD.86.052001}{\doi{10.1103/PhysRevD.86.052001}},
  \href{http://www.arXiv.org/abs/1204.3094}{\texttt{arXiv:1204.3094}}.
[Erratum: \DOI{10.1103/PhysRevD.90.079902}].

\bibitem{IceCube_1}
\hrefCMSnoop {}{{IceCube} Collaboration, ``{Search for dark matter
  annihilations in the Sun with the 79-string IceCube detector}'',} \textit{
  Phys. Rev. Lett.} \textbf{ 110} (2013) 131302,
  \href{http://dx.doi.org/10.1103/PhysRevLett.110.131302}{\doi{10.1103/PhysRevLett.110.131302}},
\href{http://www.arXiv.org/abs/1212.4097}{\texttt{arXiv:1212.4097}}.

\bibitem{SUPERK}
\hrefCMSnoop {}{{Super-Kamiokande} Collaboration, ``An indirect search for
  weakly interacting massive particles in the sun using 3109.6 days of
  upward-going muons in {Super-Kamiokande}'',} \textit{ Astrophys. J.} \textbf{
  742} (2011) 78,
  \href{http://dx.doi.org/10.1088/0004-637X/742/2/78}{\doi{10.1088/0004-637X/742/2/78}},
  \href{http://www.arXiv.org/abs/1108.3384}{\texttt{arXiv:1108.3384}}.

\bibitem{LUX}
\hrefCMSnoop {}{{LUX} Collaboration, ``{First results from the LUX dark matter
  experiment at the Sanford Underground Research Facility}'',} \textit{ Phys.
  Rev. Lett.} \textbf{ 112} (2014) 091303,
  \href{http://dx.doi.org/10.1103/PhysRevLett.112.091303}{\doi{10.1103/PhysRevLett.112.091303}},
\href{http://www.arXiv.org/abs/1310.8214}{\texttt{arXiv:1310.8214}}.

\bibitem{CDMSLITE}
\hrefCMSnoop {}{{SuperCDMS Soudan} Collaboration, ``{Search for Low-Mass Weakly
  Interacting Massive Particles Using Voltage-Assisted Calorimetric Ionization
  Detection in the SuperCDMS Experiment}'',} \textit{ Phys. Rev. Lett.}
  \textbf{ 112} (2014) 041302,
  \href{http://dx.doi.org/10.1103/PhysRevLett.112.041302}{\doi{10.1103/PhysRevLett.112.041302}},
\href{http://www.arXiv.org/abs/1309.3259}{\texttt{arXiv:1309.3259}}.

\bibitem{Agnese:2013rvf}
\hrefCMSnoop {}{{CDMS} Collaboration, ``{Silicon Detector Dark Matter Results
  from the Final Exposure of CDMS II}'',} \textit{ Phys. Rev. Lett.} \textbf{
  111} (2013) 251301,
  \href{http://dx.doi.org/10.1103/PhysRevLett.111.251301}{\doi{10.1103/PhysRevLett.111.251301}},
\href{http://www.arXiv.org/abs/1304.4279}{\texttt{arXiv:1304.4279}}.

\bibitem{atlas2015}
\hrefCMSnoop {}{{ATLAS Collaboration}, ``{Search for new phenomena in events
  with a photon and missing transverse momentum in $pp$ collisions at
  $\sqrt{s}=8$ TeV with the ATLAS detector}'',} \textit{ Phys. Rev. D} \textbf{
  91} (2015) 012008,
  \href{http://dx.doi.org/10.1103/PhysRevD.91.012008}{\doi{10.1103/PhysRevD.91.012008}},
\href{http://www.arXiv.org/abs/1411.1559}{\texttt{arXiv:1411.1559}}.

\bibitem{CDFgamma}
\hrefCMSnoop {}{{CDF} Collaboration, ``{Search for large extra dimensions in
  final states containing one photon or jet and large missing transverse energy
  produced in p$\bar{\text{p}}$ collisions at $\sqrt{s}= 1.96$~TeV}'',}
  \textit{ Phys. Rev. Lett.} \textbf{ 101} (2008) 181602,
  \href{http://dx.doi.org/10.1103/PhysRevLett.101.181602}{\doi{10.1103/PhysRevLett.101.181602}},
\href{http://www.arXiv.org/abs/0807.3132}{\texttt{arXiv:0807.3132}}.

\bibitem{D0gamma}
\hrefCMSnoop {}{{D0} Collaboration, ``{Search for Large Extra Dimensions via
  Single Photon plus Missing Energy Final States at $\sqrt{s} = 1.96$ TeV}'',}
  \textit{ Phys. Rev. Lett.} \textbf{ 101} (2008) 011601,
  \href{http://dx.doi.org/10.1103/PhysRevLett.101.011601}{\doi{10.1103/PhysRevLett.101.011601}},
  \href{http://www.arXiv.org/abs/0803.2137}{\texttt{arXiv:0803.2137}}.

\bibitem{LEPgamma}
\hrefCMSnoop {}{{DELPHI} Collaboration, ``{Photon events with missing energy in
  e$^+$ e$^-$ collisions at $\sqrt{s}= 130$~GeV to $209$~GeV}'',} \textit{ Eur.
  Phys. J. C} \textbf{ 38} (2005) 395,
  \href{http://dx.doi.org/10.1140/epjc/s2004-02051-8}{\doi{10.1140/epjc/s2004-02051-8}},
  \href{http://www.arXiv.org/abs/hep-ex/0406019}{\texttt{arXiv:hep-ex/0406019}}.

\bibitem{L3gamma}
\hrefCMSnoop {}{{L3} Collaboration, ``{Single- and multi-photon events with
  missing energy in e$^{+}$ e$^{-}$ collisions at LEP}'',} \textit{ Phys. Lett.
  B} \textbf{ 587} (2004) 16,
  \href{http://dx.doi.org/10.1016/j.physletb.2004.01.010}{\doi{10.1016/j.physletb.2004.01.010}},
\href{http://www.arXiv.org/abs/hep-ex/0402002}{\texttt{arXiv:hep-ex/0402002}}.

\bibitem{OPAL}
\hrefCMSnoop {}{{OPAL} Collaboration, ``{Photonic events with missing energy in
  e$^{+}$ e$^{-}$ collisions at $\sqrt{s}$ = 189 GeV}'',} \textit{ Eur. Phys.
  J. C} \textbf{ 18} (2000) 253,
  \href{http://dx.doi.org/10.1007/s100520000522}{\doi{10.1007/s100520000522}},
\href{http://www.arXiv.org/abs/hep-ex/0005002}{\texttt{arXiv:hep-ex/0005002}}.

\bibitem{ALEPH}
\hrefCMSnoop {}{{ALEPH} Collaboration, ``{Single photon and multiphoton
  production in e$^{+}$ e$^{-}$ collisions at $\sqrt{s}$ up to 209 GeV}'',}
  \textit{ Eur. Phys. J. C} \textbf{ 28} (2003) 1,
\href{http://dx.doi.org/10.1140/epjc/s2002-01129-7}{\doi{10.1140/epjc/s2002-01129-7}}.

\bibitem{monop}
\hrefCMSnoop {}{{CMS Collaboration}, ``{Search for Dark Matter and Large Extra
  Dimensions in $\Pp\Pp$ Collisions Yielding a Photon and Missing Transverse
  Energy}'',} \textit{ Phys. Rev. Lett.} \textbf{ 108} (2012) 261803,
  \href{http://dx.doi.org/10.1103/PhysRevLett.108.261803}{\doi{10.1103/PhysRevLett.108.261803}},
\href{http://www.arXiv.org/abs/1204.0821}{\texttt{arXiv:1204.0821}}.

\bibitem{An:2012va}
\hrefCMSnoop {}{H.~An, X.~Ji, and L.-T. Wang, ``{Light dark matter and $Z'$
  dark force at colliders}'',} \textit{ JHEP} \textbf{ 07} (2012) 182,
  \href{http://dx.doi.org/10.1007/JHEP07(2012)182}{\doi{10.1007/JHEP07(2012)182}},
\href{http://www.arXiv.org/abs/1202.2894}{\texttt{arXiv:1202.2894}}.

\bibitem{Friedland:2011za}
\hrefCMSnoop {}{A.~Friedland, M.~L. Graesser, I.~M. Shoemaker, and L.~Vecchi,
  ``{Probing nonstandard standard model backgrounds with LHC monojets}'',}
  \textit{ Phys. Lett. B} \textbf{ 714} (2012) 267,
  \href{http://dx.doi.org/10.1016/j.physletb.2012.06.078}{\doi{10.1016/j.physletb.2012.06.078}},
\href{http://www.arXiv.org/abs/1111.5331}{\texttt{arXiv:1111.5331}}.

\bibitem{Buchmueller:2013dya}
\hrefCMSnoop {}{O.~Buchmueller, M.~J. Dolan, and C.~McCabe, ``Beyond effective
  field theory for dark matter searches at the {LHC}'',} \textit{ JHEP}
  \textbf{ 01} (2014) 025,
  \href{http://dx.doi.org/10.1007/JHEP01(2014)025}{\doi{10.1007/JHEP01(2014)025}},
\href{http://www.arXiv.org/abs/1308.6799}{\texttt{arXiv:1308.6799}}.

\bibitem{l3}
\hrefCMSnoop {}{{L3} Collaboration, ``{Search for branons at LEP}'',} \textit{
  Phys. Lett. B} \textbf{ 597} (2004) 145,
  \href{http://dx.doi.org/10.1016/j.physletb.2004.07.014}{\doi{10.1016/j.physletb.2004.07.014}},
\href{http://www.arXiv.org/abs/hep-ex/0407017}{\texttt{arXiv:hep-ex/0407017}}.

\bibitem{eft_add}
\hrefCMSnoop {}{G.~F. Giudice, R.~Rattazzi, and J.~D. Wells, ``{Quantum gravity
  and extra dimensions at high-energy colliders}'',} \textit{ Nucl. Phys. B}
  \textbf{ 544} (1999) 3,
  \href{http://dx.doi.org/10.1016/S0550-3213(99)00044-9}{\doi{10.1016/S0550-3213(99)00044-9}},
\href{http://www.arXiv.org/abs/hep-ph/9811291}{\texttt{arXiv:hep-ph/9811291}}.

\bibitem{Aad:2012fw}
\hrefCMSnoop {}{{ATLAS Collaboration}, ``{Search for dark matter candidates and
  large extra dimensions in events with a photon and missing transverse
  momentum in $pp$ collision data at $\sqrt{s}=7$ TeV with the ATLAS
  detector}'',} \textit{ Phys. Rev. Lett.} \textbf{ 110} (2013) 011802,
  \href{http://dx.doi.org/10.1103/PhysRevLett.110.011802}{\doi{10.1103/PhysRevLett.110.011802}},
\href{http://www.arXiv.org/abs/1209.4625}{\texttt{arXiv:1209.4625}}.

\end{thebibliography}\endgroup
